\documentclass{JHEP3}
\usepackage{graphicx}
\usepackage{latexsym}
\usepackage{epsfig}
\usepackage{amsmath}
\usepackage{amsfonts}
\usepackage{amssymb}  
\usepackage{mathrsfs}
\usepackage{epsfig,bbm}
\usepackage{enumerate}

\newcommand\fverb{\setbox\fverbbox=\hbox\bgroup\verb}
\newcommand\fverbdo{\egroup\medskip\noindent%
\fbox{\unhbox\fverbbox}\ }
\newcommand\fverbit{\egroup\item[\fbox{\unhbox\fverbbox}]}
\newbox\fverbbox

\newcommand{\data}{d}
\newcommand{\nuis}{\psi}
\newcommand{\params}{\theta}
\newcommand{\basis}{\Theta}

\newcommand\etal{{\it {et al.}}}

\newcommand{\be}{\begin{equation}}
\newcommand{\ee}{\end{equation}}
\newcommand{\beq}{\begin{equation}}
\newcommand{\eeq}{\end{equation}}
\newcommand{\bea}{\begin{eqnarray}}
\newcommand{\eea}{\end{eqnarray}}
\newcommand{\gsim}{\lower.7ex\hbox{$\;\stackrel{\textstyle>}{\sim}\;$}}
\newcommand{\lsim}{\lower.7ex\hbox{$\;\stackrel{\textstyle<}{\sim}\;$}}

\newcommand{\mhalf}{m_{1/2}}      \newcommand{\mzero}{m_0}
\newcommand{\tanb}{\tan\beta}
\newcommand{\azero}{A_0}

\newcommand{\cl}{\text{CL}}
\newcommand{\mtpole}{M_t}
\newcommand{\alphas}{\alpha_s(M_Z)^{\overline{MS}}}
\newcommand{\alphaemmz}{\alpha_{\text{em}}(M_Z)^{\overline{MS}}}

\newcommand{\BR}{BR}

\newcommand{\brbsgamma}{\BR(\overline{B}\rightarrow X_s\gamma)}

\newcommand\delmbs{\Delta M_{B_s}}

\newcommand\brbsmumu{\BR(\overline{B}_s\to\mu^+\mu^-)}

\newcommand\RBtaunu{\frac{\BR(B_u \to \tau \nu)}{\BR(B_u \to \tau \nu)_{SM}}}
\newcommand\DeltaO{\Delta_{0-}}
\newcommand\RBDtaunuBDenu{\frac{\BR(B \to D \tau \nu)}{\BR(B \to D e \nu)}}
\newcommand\Rl{R_{l23}}

\newcommand\Dstaunu{\BR(D_s \to \tau \nu)}
\newcommand\Dsmunu{\BR(D_s\to \mu \nu)} 
\newcommand\Dmunu{\BR(D \to \mu \nu)}

\newcommand{\gev}{\mbox{ GeV}}
\newcommand{\tev}{\mbox{ TeV}}

\newcommand{\mhl}{m_h}
\newcommand{\zetah}{\zeta_h}

\newcommand{\mchi}{m_\chi}

\newcommand{\rholoc}{\rho_{\odot}}

\newcommand{\abundchi}{\Omega_{\chi}\,h^2}

\newcommand{\sigsip}{\sigma_{\chi{\rm p}}^\text{SI}}

\newcommand{\ETslash}{/ \hspace{-.7em} E_T}

\title{Dark Matter Searches: The Nightmare Scenario} 
 \author{Gianfranco Bertone${}^{1,2}$, Daniel Cumberbatch${}^3$, Roberto Ruiz de Austri${}^4$ and Roberto Trotta${}^{5,6}$\\ 
 ${}^1$Institute for Theoretical Physics, Univ. of Z\"urich, Winterthurerst. 190, 8057 Z\"urich, CH\\ 
 ${}^2$Institut d'Astrophysique de Paris, UMR 7095-CNRS, Univ. P. et M. Curie, 98bis Bd Arago, 75014 Paris, France\\ 
 ${}^3$Astroparticle Theory and Cosmology Group, University of Sheffield, Hicks Building, Hounsfield Road, Sheffield, South Yorkshire, S3 7RH, UK\\ 
 ${}^4$Instituto de F\'isica Corpuscular, IFIC-UV/CSIC, Apartado 22085, E-46071 Valencia, Spain\\ 
 ${}^5$Astrophysics Group, Imperial College London, Blackett Laboratory, Prince Consort Road, London SW7 2AZ, UK\\ 
 ${}^6$Kavli Institute for Theoretical Physics, Kohn Hall, University of California, Santa Barbara, CA 93106-4030, USA\\
 Email: \email{bertone@iap.fr}, \email{d.cumberbatch@sheffield.ac.uk}, \email{rruiz@ific.uv.es}, \email{r.trotta@imperial.ac.uk}}
\abstract{The unfortunate case where the Large Hadron Collider (LHC) fails to 
discover physics Beyond the Standard Model (BSM) is sometimes referred to as 
the ``Nightmare scenario'' of particle physics. We study the consequences of this 
hypothetical scenario for Dark Matter (DM), in the framework of the constrained 
Minimal Supersymmetric Standard Model (cMSSM). We evaluate the surviving 
regions of the cMSSM parameter space after null searches at the LHC, using 
several different LHC configurations, and study the consequences for DM 
searches with ton-scale direct detectors and the IceCube neutrino telescope.  
We demonstrate that ton-scale direct detection experiments will be able 
to conclusively probe the cMSSM parameter space that would survive null searches at the LHC with 100\,fb$^{-1}$ of integrated luminosity 
at 14\,TeV.
We also demonstrate that IceCube (80 strings plus DeepCore) will be able to probe as much as 
$\simeq17\%$ of the currently favoured parameter space after 5\,years of observation.}
\begin{document}

%%%%%%%%%%%%%%%%%%%%%%%%%%%%%%%%%%%%%%%
\section{Introduction}
\label{sec:intro}
%%%%%%%%%%%%%%%%%%%%%%%%%%%%%%%%%%%%%%%

One of the main goals of the Large Hadron Collider (LHC) at CERN is to test the 
existence of physics Beyond the Standard Model (BSM), especially in view of the 
possible connection with the problem of Dark Matter (DM) in astrophysics and 
cosmology \cite{Bertone:2010zz}. Recently, the ATLAS and CMS experimental 
collaborations have published the results of the first LHC searches for Supersymmetry 
(SUSY) \cite{Chatrchyan:2011wc,Khachatryan:2011tk,daCosta:2011qk,Aad:2011hh,
Aad:2011ks,Aad:2011xm}, based on  $\sqrt{s} = 7$\,TeV collisions recorded during 
2010. The absence of any excess above Standard Model (SM) predictions allows 
one to already set interesting constraints on BSM physics, and several groups of 
authors have already studied the impact of these results on Supersymmetric 
scenarios (see, e.g., \cite{Buchmueller:2011aa,Allanach:2011ut,Akula:2011dd,arXiv:1104.3572,
Strumia:2011dv,Buchmueller:2011ki,Bertone:2011nj}).  

The prospects for discovering BSM physics at the LHC and the consequences for 
Dark Matter searches have been thoroughly discussed in the literature. In particular, 
we have recently discussed the case of positive detection both at the LHC {\it and} 
with Dark Matter experiments \cite{Bertone:2010rv}. Here, we focus instead on the 
unfortunate case where the LHC {\it fails} to discover BSM physics, a scenario 
sometimes referred to as the ``Nightmare scenario'' of particle physics (see, e.g., 
\cite{nature} and references therein). We focus on the constrained Minimal 
Supersymmetric Standard Model (cMSSM) \cite{CTP-TAMU-24-92,RAL-92-005,hep-ph/9311269,kkrw94,sugra-reviews}, for which 
detailed studies exist regarding the reach of the LHC for various configurations of 
beam energy and total integrated luminosity. 

The main focus of this paper is to analyse the consequences of the nightmare scenario 
for Dark Matter searches. These can be actually divided into two broad classes, viz. 
{\it direct} and {\it indirect} searches (for recent reviews see, e.g., \cite{Bertone:2010zz,
Bergstrom,Munoz, Bertone:2004pz,Feng:2010gw}). Direct searches are based on the search 
for rare nuclear recoils due to the scattering of Dark Matter particles off nuclei in 
underground experiments. There are many undergoing and upcoming experiments
(see, e.g., \cite{Pato:2010zk} for a discussion of upcoming experimental capabilities), 
and despite some intriguing signals that have been discovered by the DAMA/LIBRA 
\cite{dama} and, more recently, the CoGeNT \cite{cogent,Aalseth:2011wp,Hooper:2010uy} 
collaborations, it appears difficult to reconcile a standard Dark Matter interpretation of 
these results with other experimental findings, in particular with the null searches from 
XENON-100 \cite{xe100} or CDMS II \cite{cdms10} (see, e.g., the discussions in 
\cite{gelmini1,Bezrukov,Arina:2011si,Schwetz:2011xm}). 

Indirect searches are based instead on the astrophysical searches for secondary particles 
produced in the annihilation or decay of Dark Matter particles. Despite the huge interest 
attracted by astrophysical data (see, in particular, the plethora of literature devoted to 
explaining the positron excess recently measured by PAMELA  \cite{Adriani:2008zr}), 
a convincing identification of Dark Matter based on these data seems problematic, due 
to the lack of strong constraints on the properties of Dark Matter particles as well as to 
the poor control of astrophysical backgrounds and associated systematics (see, e.g., 
the discussion in \cite{nature}). Among the possible smoking-gun signals that can be 
obtained with indirect searches, we will focus on one of the most intriguing, and less 
affected by astrophysical uncertainties: the detection of high-energy neutrinos from the 
centre of the Sun \cite{Silk:1985ax} using the IceCube neutrino telescope, which is located 
at the geographical South Pole (see, e.g., \cite{Halzen:2009vu}).

We show here that upcoming ton-scale direct detection (DD) experiments, and to a lesser 
extent IceCube, can actually probe a large portion of cMSSM parameter space that will 
remain unconstrained in the nightmare scenario, and therefore represent {\it a unique 
opportunity to discover new physics by the end of the decade}, in case of null searches 
at the LHC.  

The paper is organized as follows: in Sec.\,\ref{sec:cmssm} we introduce our theoretical 
framework, i.e., the cMSSM,  and we present the statistical tools adopted to identify the 
region of cMSSM parameter space corresponding to the LHC {\it nightmare} scenario; in 
Sec.\,\ref{sec:results} we present our results and discuss the consequences for Dark Matter 
searches; and, finally, in Sec.\,\ref{sec:summary} we provide a brief summary.

%%%%%%%%%%%%%%%%%%%%%%%%%%%%%%%%%%%%%%%
\section{Theoretical Setup}
\label{sec:cmssm}
%%%%%%%%%%%%%%%%%%%%%%%%%%%%%%%%%%%%%%%								

%%%%%%%%%%%%%%%%%%%%%%%%%%%%%%%%%%%%%%%
\subsection{The cMSSM}
%%%%%%%%%%%%%%%%%%%%%%%%%%%%%%%%%%%%%%%

Supersymmetry (SUSY) is  one the most compelling extensions of the Standard Model 
of Particle Physics, and the Supersymmetric neutralino is among the best-motivated
Dark Matter (DM) candidates (for a review of SUSY DM candidates see, e.g.,\,
\cite{Jungman:1995df, Bertone:2004pz, Feng:2010gw}.)

Neutralinos are the mass eigenstate of a mixture of bino, wino (superpartners of the 
$B$ and $W^0$ gauge bosons respectively) and two higgsinos (superpartners of 
the $H^0_1$, $H^0_2$ Higgs bosons). In many realizations of the Minimal Supersymmetric 
Standard Model (MSSM), the lightest among the four neutralinos is the lightest SUSY 
particle, which is made stable by virtue of the conservation of R--parity. It is commonly 
referred to as  ``the'' neutralino, $\chi^0_1$, and
constitutes a popular DM candidate, adopted by much of the relevant literature.

Since superpartners of Standard Model particles have not been observed, SUSY 
must be broken at some energy scale. Over the past few decades, several 
SUSY--breaking mechanisms have been proposed in the literature (for a recent 
review see, e.g., \cite{Kitano:2010fa}). Perhaps the most appealing of these regards 
gravity as the means of communication between a hidden sector, where SUSY breaking 
occurs, and the visible sector \cite{sugra-reviews}.
In the  so--called constrained Minimal Supersymmetric Standard Model (cMSSM) 
\cite{CTP-TAMU-24-92,RAL-92-005,hep-ph/9311269,kkrw94,sugra-reviews}, which is the model focussed upon in this paper, the 
mechanism via which SUSY is broken nor how this breaking is transmitted to the visible 
sector need be specified, so long as the latter occurs at or near the GUT scale.
In the cMSSM the soft parameters are assumed universal at a high scale ($M_X$),
hence the parameter space of the model is defined in 
terms of four free parameters and one sign: the common scalar mass ($\mzero$), 
the common gaugino mass ($\mhalf$) and the tri--linear mass ($\azero$) parameters 
(all specified at the GUT scale), plus the ratio of Higgs vacuum expectation values 
$\tanb$ and $\text{sign}(\mu)$, where $\mu$ is the higgsino mass parameter whose 
value is determined from the conditions of radiative electroweak symmetry breaking. 

%%%%%%%%%%%%%%%%%%%%%%%%%%%%%%%%%%%%%%%
\subsection{Statistical framework}
\label{subsec:scan}
%%%%%%%%%%%%%%%%%%%%%%%%%%%%%%%%%%%%%%%

We denote the parameter set of the cMSSM (i.e., $\mzero$, $\mhalf$, $\azero$ and 
$\tanb$) by $\params$, where we fix sgn($\mu$)=+1, motivated by consistency arguments 
involving measurements of the anomalous muon magnetic moment, which can only be 
explained with a positive $\mu$ \cite{Trotta:2009gr} (see also, e.g., \cite{arXiv:0807.4512}), 
while $\nuis$ denotes the relevant SM quantities that enter into the calculation of 
observable quantities (the so--called ``nuisance parameters''), namely
%***********************************************************************************
\begin{equation}
\nuis \equiv \{ \mtpole, m_b(m_b)^{\overline{MS}}, \alphas , \alphaemmz \} \, ,
\end{equation}
%***********************************************************************************
where $\mtpole$ is the pole top quark mass, $m_b(m_b)^{\overline{MS}}$ is the 
bottom quark mass at $m_b$, while $\alphaemmz$ and $\alphas$ are the 
electromagnetic and the strong coupling constants at the $Z$ pole mass $M_Z$, 
the last three being evaluated in the $\overline{MS}$ scheme. We denote the full 
8--dimensional parameter set by
%***********************************************************************************
\be
\basis = (\params,\nuis).
\label{basis:eq}
\ee
%***********************************************************************************
The cornerstone of Bayesian inference is  Bayes' Theorem, which reads
%***********************************************************************************
\be \label{eq:bayes}
 p(\basis | \data) = \frac{p(\data |\basis) p(\basis)}{p(\data)}. 
 \ee
%***********************************************************************************
The quantity $p(\basis | \data)$ on the l.h.s.\,\,of Eq.\,\eqref{eq:bayes} is known as 
the {\em posterior}, while on the r.h.s., the quantity $p(\data |\basis)$ is the {\it likelihood} 
(when taken as a function $\basis$ for fixed data, $d$). The quantity $p(\basis)$ 
is the {\em prior} which encodes our state of knowledge about the values of the 
parameters $\basis$ before scrutinising the data. Our state of knowledge is then 
updated to the posterior via the likelihood.  Finally, the quantity in the denominator 
of Eq.\,(\ref{basis:eq}) is known as the {\em evidence} or {\em model likelihood}. If 
one is interested in constraining the model's parameters, the evidence is merely a 
normalization constant, independent of $\basis$, and can therefore be dropped 
by converting Eq.\,(\ref{eq:bayes}) into a proportionality relation (see, e.g.,~\cite{Trotta:2008qt} 
for further details).

In order to explore the posterior of Eq.\,\eqref{eq:bayes} in an efficient way, we 
adopt the \texttt{MultiNest} \cite{Feroz:2007kg} algorithm, as implemented in the 
public package~\texttt{SuperBayeS-v1.5}. \texttt{MultiNest} provides an extremely 
efficient sampler even for likelihood functions defined over a parameter space of 
large dimensionality with a very complex structure. This aspect is very important for 
exploring the cMSSM, as previous MCMC scans have revealed that the 8-dimensional 
likelihood surface is very fragmented and that it presents many finely--tuned regions 
that are difficult to explore with conventional MCMC scans (and almost impossible to 
find with conventional grid scans) (see also, e.g.,~\cite{Feroz:2010dj, Akrami:2009hp} 
for further discussions of these aspects).
  
%%%%%%%%%%%%%%%%%%%%%%%%%%%%%%%%%%%%%%%
\subsection{Priors} 
\label{subsec:prior}
%%%%%%%%%%%%%%%%%%%%%%%%%%%%%%%%%%%%%%%

In order to perform our scan over the cMSSM and SM nuisance parameters, we 
need to specify our prior in Eq.\,\eqref{eq:bayes}. The role of the prior is to define 
a statistical measure on the parameter space. In principle, when the likelihood is 
strongly constraining (i.e., for accurate data) the posterior is dominated by the 
likelihood and the choice of prior is irrelevant, as the information in the likelihood 
completely overrides the information in the prior. However, it has been shown that 
this is presently not the case for the cMSSM, i.e., different, plausible, choices of 
priors lead to different posteriors and hence different inferences on cMSSM parameter 
space~\cite{tfhrr:2008}, although inclusion of recent direct detection constraints 
from XENON-100 does mitigate the problem~\cite{Bertone:2011nj}. It is expected 
that a complete resolution of this issue will come from future, more detailed, data 
and, in particular, measurements of the SUSY mass spectrum by the LHC, which 
will conclusively resolve ambiguities brought about by prior dependences in case 
SUSY is discovered~\cite{Roszkowski:2009ye}. 

In this paper, we adopt two sets of widely used priors, namely the so-called ``flat 
priors'' (uniform on the scalar and gaugino masses) and the ``log-priors'' (uniform 
on the log of the masses). Both sets of priors are uniform in $A_0$ and $\tan\beta$. 
By considering two choices of priors, we can assess the residual prior dependency 
of our conclusions. The prior ranges we use are given by
%***********************************************************************************
\begin{eqnarray}
& 50 \,{\rm GeV} \leq  m_0, \mhalf \leq 4000 \,{\rm GeV} \nonumber\\
& |A_0| \leq7\tev \nonumber \\
& 2 < \tan\beta < 62. \nonumber
\end{eqnarray}
 %***********************************************************************************

 %***********************************************************************************
\begin{table}[t]
\centering
\begin{tabular}{|l | l l l | l|}
\hline
Observable &   Mean value & \multicolumn{2}{c|}{Uncertainties} & ref. \\
 &   $\mu$      & ${\sigma}$ (exper.)  & $\tau$ (theor.) & \\\hline
$M_W$ [GeV] & 80.398 & 0.025 & 0.015 & \cite{lepwwg} \\
$\sin^2\theta_{eff}$ & 0.23153 & 0.00016 & 0.00015 & \cite{lepwwg} \\
$\delta a_\mu^{SUSY}\times10^{10}$ & 29.6 & 8.1 & 2.0 & \cite{Davier:tau10} \\
$BR(\bar{B} \rightarrow X_s\gamma)\times10^4$ & 3.55 & 0.26 & 0.30 & \cite{hfag}\\
$\Delta M_{B_s}$ [ps$^{-1}$] & 17.77 & 0.12 & 2.40 & \cite{cdf-deltambs} \\
$\RBtaunu$   &  1.28  & 0.38  & - & \cite{hfag}  \\
$\DeltaO  \times 10^{2}$   &  3.6  & 2.65  & - & \cite{:2008cy}  \\
$\RBDtaunuBDenu \times 10^{2}$ & 41.6 & 12.8 & 3.5  & \cite{Aubert:2007dsa}  \\
$\Rl$ & 1.004 & 0.007 & -  &  \cite{Antonelli:2008jg}  \\
$\Dstaunu\times10^{2}$ & 5.38 & 0.32 & 0.2  & \cite{hfag}  \\
$\Dsmunu\times10^{3}$ & 5.81 & 0.43 & 0.2  & \cite{hfag}  \\
$\Dmunu \times10^{4}$  & 3.82  & 0.33 & 0.2  & \cite{hfag} \\
$\Omega_\chi h^2$ & 0.1123 & 0.0035 & 10\% & \cite{Jarosik:2010iu} \\\hline\hline
 &  \multicolumn{2}{l}{Limit (95\%~\cl)}  & $\tau$ (theor.) & ref. \\ \hline
$\brbsmumu$ &  $ <5.8\times 10^{-8}$  & & 14\%  & \cite{cdf-bsmumu}\\
$\mhl$  & \multicolumn{2}{l}{$>114.4\gev$}  & $3 \gev$ & \cite{lhwg} \\
$\zetah^2$ &  \multicolumn{3}{l |}{$f(m_h)$ as defined in~\cite{deAustri:2006pe}} & \cite{lhwg} \\
$m_{\tilde{q}}$   &  $>375$\gev     &  & 5\%                     & \cite{pdg07}\\
$m_{\tilde{g}}$   &  $>289$\gev     &  & 5\%                     & \cite{pdg07}\\
other sparticle &  \multicolumn{3}{l|}{As in Table\,4 of \cite{deAustri:2006pe}.} 
& \cite{deAustri:2006pe} \\ 
masses          &                  &   &                         &  \\
\hline
\end{tabular}
\caption{\fontsize{9}{9} \selectfont Experimental data used for the computation of the 
likelihood function. For each row, the central value is given, together with the experimental 
and theoretical uncertainty.}\label{tab:obs}
\end{table}
 %***********************************************************************************

 %***********************************************************************************
\begin{table}[t]
\centering
\begin{tabular}{|l | l l | l|}
\hline
SM (nuisance)                             & Mean value            & Uncertainty            & Ref. \\
parameter                                    & $\mu$                       & ${\sigma}$ (exp.)  &         \\ 
\hline
$\mtpole$                                     & 173.1\gev                & 1.3\gev                   & \cite{topmass:mar08} \\
$m_b (m_b)^{\overline{MS}}$  & 4.20\gev                  & 0.07\gev                 & \cite{pdg07} \\
$\alphas$                                     & 0.1176                     & $2\times10^{-3}$  & \cite{pdg07} \\
$1/\alphaemmz$                         & 127.955                  & 0.03                         & \cite{Hagiwara:2006jt} \\ 
\hline
\end{tabular}
\caption{Experimental mean $\mu$ and standard deviation $\sigma$ 
 adopted for the likelihood function for SM (nuisance) parameters,
 assumed to be described by a Gaussian probability distribution function.
\label{tab:nuisance}}
\end{table}
 %***********************************************************************************

%%%%%%%%%%%%%%%%%%%%%%%%%%%%%%%%%%%%%%%
\subsection{Cosmological and collider constraints}
\label{subsec:observ}
%%%%%%%%%%%%%%%%%%%%%%%%%%%%%%%%%%%%%%%

In order to derive current constraints on cMSSM parameters, we perform a global 
scan for both approaches using flat priors on the model parameters, including in 
the likelihood current cosmological and collider experimental constraints, as 
displayed in Table\,\ref{tab:obs}, and with constraints on the SM nuisance parameters 
as displayed in Table\,\ref{tab:nuisance}. In this paper, we utilise the same likelihood 
function as in~\cite{Bertone:2011nj}, invoking the experimental constraints listed 
in Table\,\ref{tab:obs}. We refer the reader to \cite{deAustri:2006pe} for a detailed 
description of the likelihood function. In particular, we include the measurement 
of the anomalous magnetic moment of the muon based on $e^+e^-$ data, which 
gives a $3.7\,\sigma$ discrepancy with the SM predicted value~\cite{Davier:tau10}. 

We compute $\delta_{\rm had}^{\rm SM}a_\mu$ at full one-loop level adding the 
logarithmic piece of the quantum electro-dynamics two-loop 
calculation~\cite{Degrassi:1998es}.
The $\brbsgamma$ branching ratio (which has been shown to provide 
an important constraint, see, e.g., the recent study~\cite{Roszkowski:2007fd}), 
has been computed with the 
numerical code \texttt{SusyBSG} \cite{Degrassi:2007kj} using the full NLO QCD 
contributions, including the two-loop calculation of the gluino contributions presented 
in \cite{Degrassi:2006eh} and the results of \cite{D'Ambrosio:2002ex} for the remaining 
non-QCD $\tan\beta$-enhanced contributions. For the determination of $\delmbs$ 
we use expressions from \cite{for1} which include dominant large $\tanb$-enhanced 
beyond-LO SUSY contributions from Higgs penguin diagrams. The other B(D)-physics 
observables summarized in Table\,\ref{tab:obs} have been computed with the code 
\texttt{SuperIso} (for details on the computation of the observables see, e.g., 
\cite{Mahmoudi:2008tp} and references therein). Both codes have been integrated 
into \texttt{SuperBayes}. We discard points that do not fulfill the conditions of radiative 
electroweak symmetry breaking and/or give non-physical (i.e., tachyonic) solutions.

We also include the constraints on the cold Dark Matter (CDM) relic abundance 
determined from the 7-year WMAP data \cite{Jarosik:2010iu} to constrain the relic 
abundance $\abundchi$ of the lightest neutralino, which we assume represents the 
sole constituent of CDM in the Universe, by invoking a conventional Gaussian pdf in 
the likelihood (with an additional 10\% theoretical error).

The constraints imposed by the recent LHC data are described in detail in 
Sec.\,\ref{subsec:lhc_reach}.

%%%%%%%%%%%%%%%%%%%%%%%%%%%%%%%%%%%%%%%
\subsection{LHC constraints}
\label{subsec:lhc_reach}
%%%%%%%%%%%%%%%%%%%%%%%%%%%%%%%%%%%%%%%

The LHC at CERN started operations in September 2008, marking the beginning 
of an intensive period of investigation. The current plan (as of Summer 2011) is 
that the LHC will be running at the current  centre of mass energy $\sqrt{s}=7$\,TeV 
throughout 2012. Following this period, an approximately 18 month maintenance 
period is expected to commence in order to upgrade the LHC, after which the first 
attempts at collisions with $\sqrt{s}=14\,$TeV are expected. 

Whilst the discovery of SUSY would obviously be of paramount importance, the 
task of actually identifying weakly--interacting massive particles (WIMPs) as
the dominant constituent of DM using LHC data alone is challenging \cite{Baltz:2006fm},
unless complementary information is provided by other experiments, e.g., by 
direct~\cite{Bertone:2010rv} or indirect~\cite{Bertone:2011pq} searches, or by specific accelerator signatures, such 
as the shape of the dilepton invariant mass spectrum in the MSSM focus point region 
\cite{White:2010jp}.

To compute the region of the cMSSM parameter space where the LHC will be able 
to achieve a 5$\sigma$ discovery, we consider studies conducted by the ATLAS 
collaboration for a centre of mass energy of 14\,TeV and an integrated luminosity 
(IL) of 1\,fb$^{-1}$ \cite{atl:2010035} (which is expected to be achieved by 2014--2015)
%%%%%%%%%%%%%%%%%%%%%%%%%%%%%%%%%%%%%%%%%%%
\footnote{We have also investigated the LHC configuration with $\sqrt{s}=$7\,GeV and 
IL\,=\,1\,fb$^{-1}$, which is very close to what has already been achieved \cite{ATLASsusy, ATLASHA, CMSsusy},
and consequently, we omit our results for this case.}. 
%%%%%%%%%%%%%%%%%%%%%%%%%%%%%%%%%%%%%%%%%%%
We have adopted the analysis consisting in looking for events containing 4 jets + 0 leptons + $\ETslash$ which have 
been performed for fixed $\azero=0, \tanb=10$ values. Nevertheless these sort 
of analyses are relatively insensitive to $\tanb$ and $\azero$ due to the fact that 
large $\tanb$ values, basically, alter the phenomenology of multilepton production 
channels \cite{Baer:1997yi} which are not considered here. We have followed the 
procedure described in \cite{Baer:2009dn} for implementing the $\sqrt{s}=$14\,TeV and 
IL\,=\,100\,fb$^{-1}$ LHC configuration (representing what might be achieved over a 
timescale of about 10 years). In this case, the authors have optimized the search 
looking for events containing a number of jets $>2$ + $n=0,..,6$ leptons + $\ETslash$ 
where  $\azero=0$, $\tanb=45$ values have been assumed. Studies of this sort have 
shown that the detectability level exhibits only a mild dependence on $\azero$ and 
$\tanb$~\cite{Baer:1998sz}. Therefore, although the resulting LHC sensitivity is 
expressed as a detection region in the $\mzero$, $\mhalf$ plane for fixed values of 
$\azero$ and $\tanb$, we assume it to be universal for all values of $\azero$ and 
$\tanb$, as argued above.

 %***********************************************************************************
\begin{center}
\begin{figure*}[t]
\includegraphics[width=0.32\textwidth,keepaspectratio,clip]{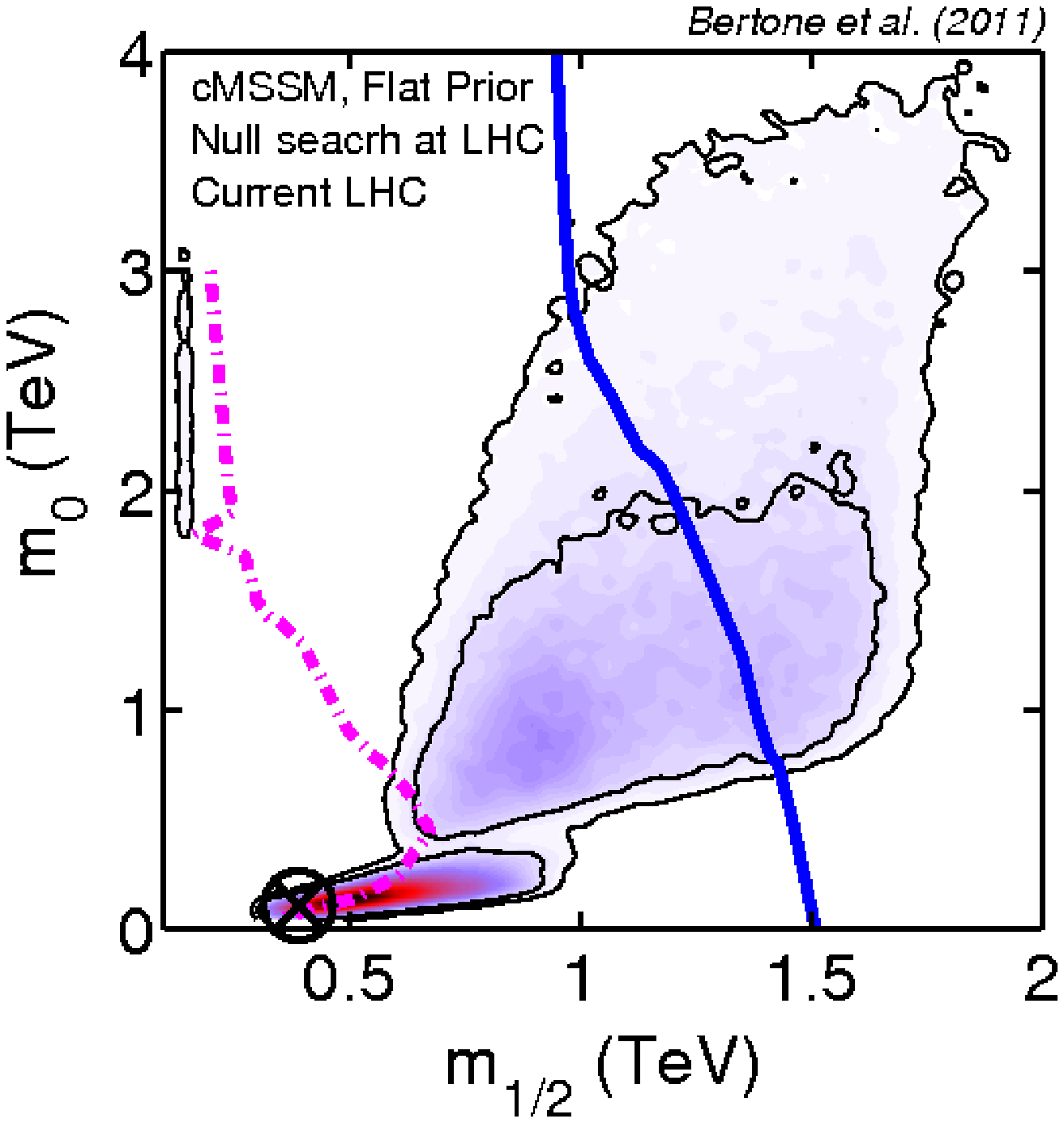}
\includegraphics[width=0.32\textwidth,keepaspectratio,clip]{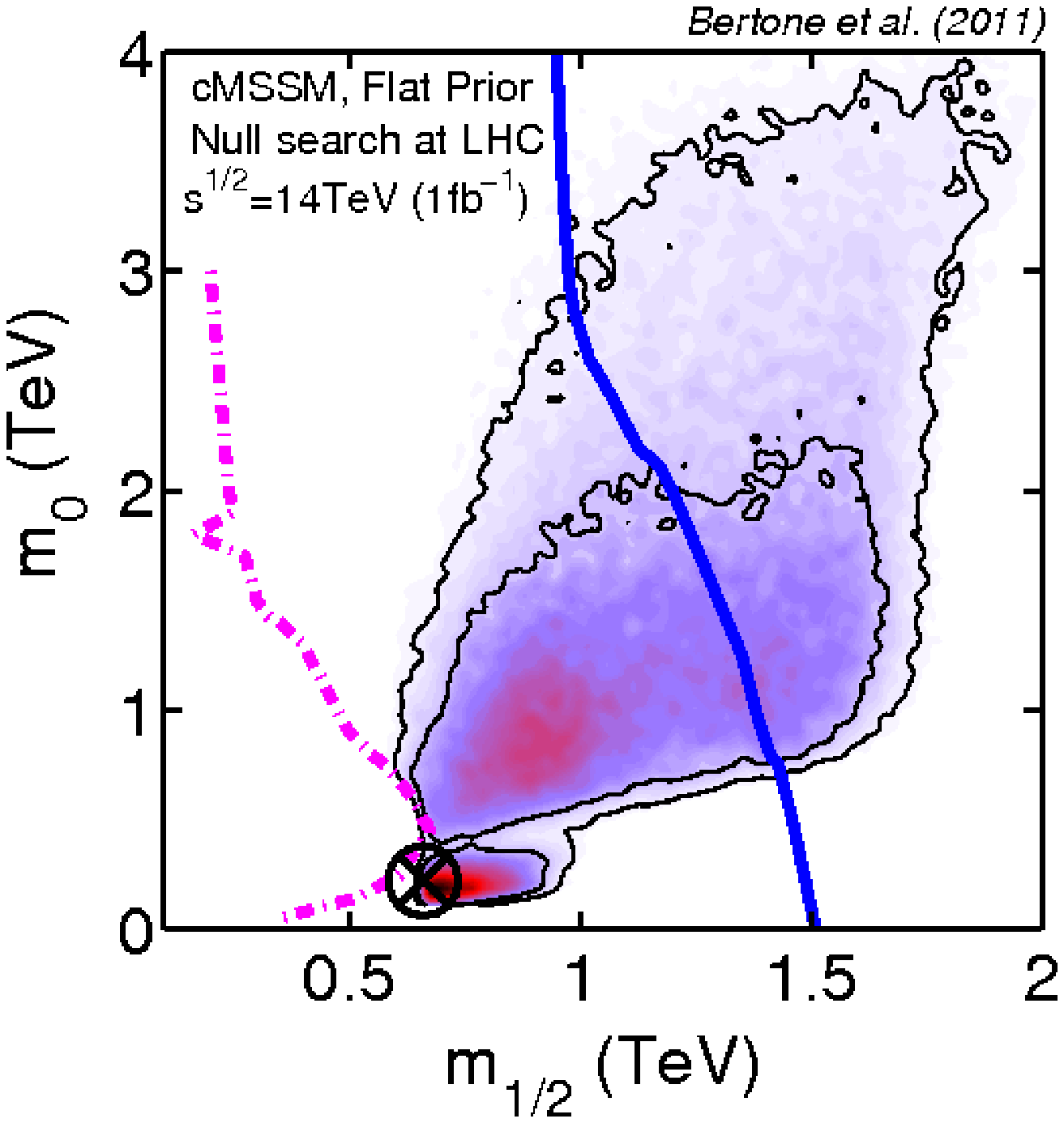}
\includegraphics[width=0.32\textwidth,keepaspectratio,clip]{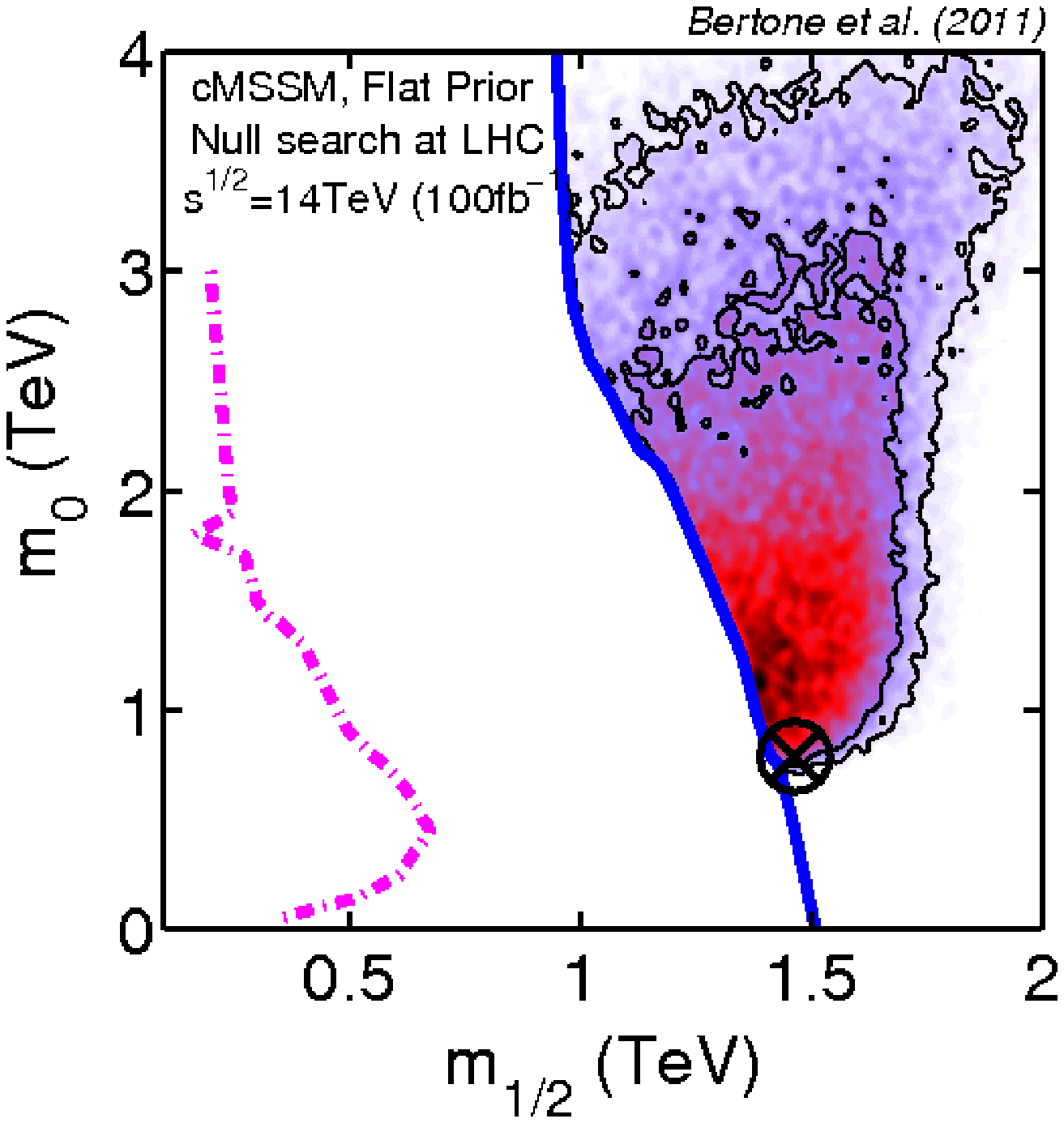}\\
\includegraphics[width=0.32\textwidth,keepaspectratio,clip]{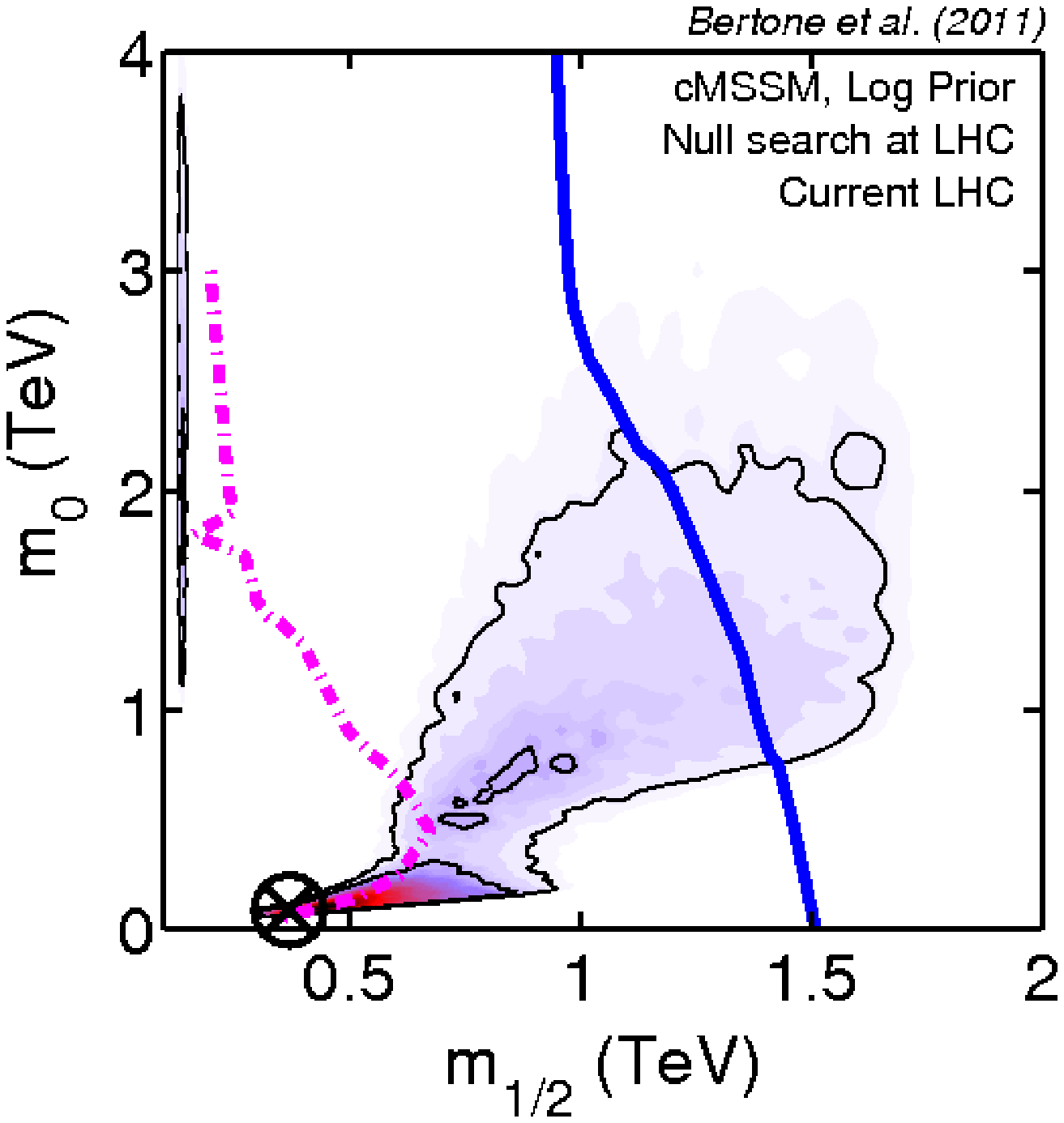}
\includegraphics[width=0.32\textwidth,keepaspectratio,clip]{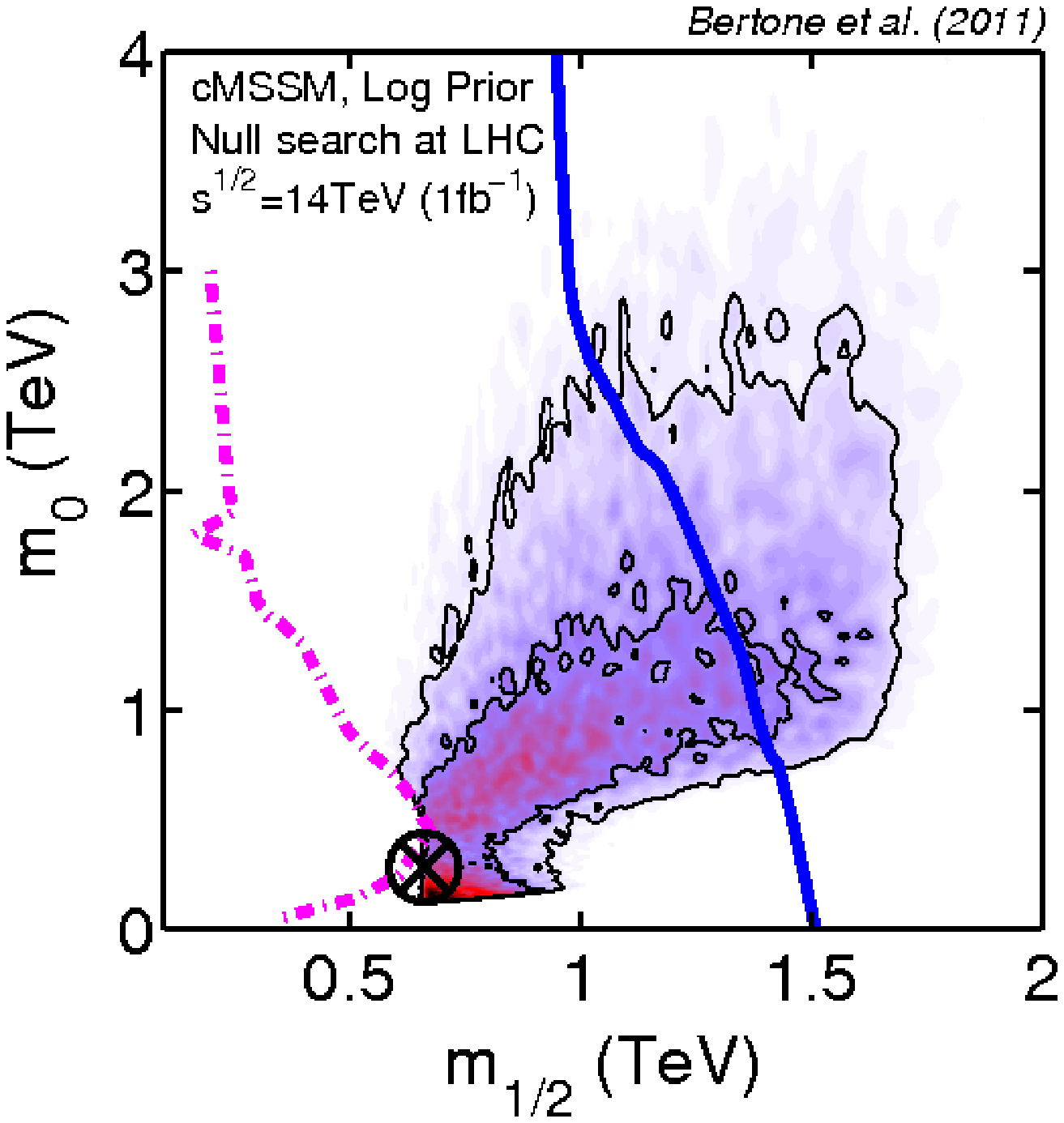}
\includegraphics[width=0.32\textwidth,keepaspectratio,clip]{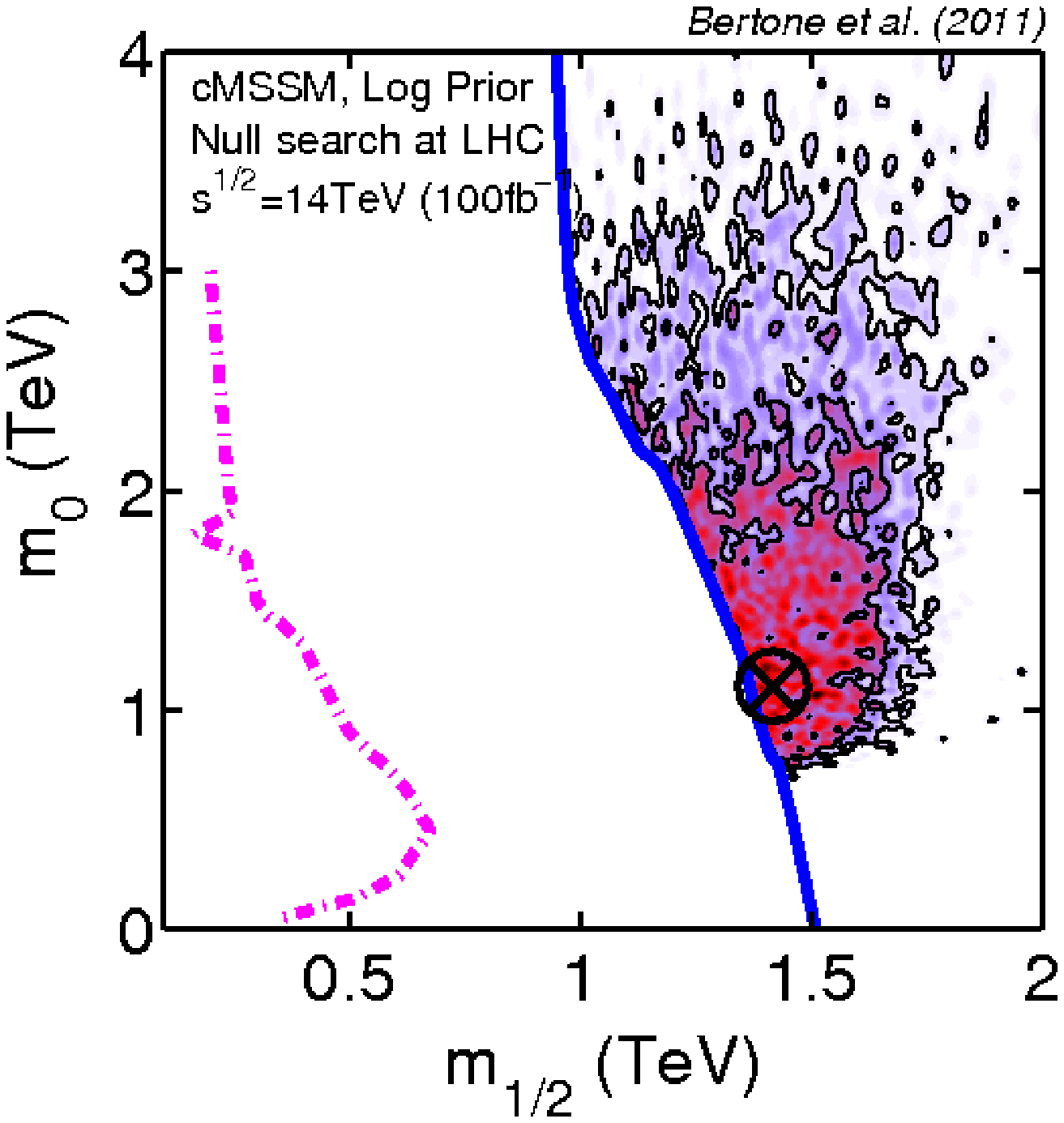}
\caption{Posterior probability distribution for the cMSSM in the ($m_0$, $m_{1/2}$) 
plane, after null searches by the LHC with combinations of $\sqrt{s}$ and integrated 
luminosities (IL) (left to right): Current LHC, 14\,TeV and 1\,fb$^{-1}$ and 14\,TeV and 
100\,fb$^{-1}$, for the priors (top) and log priors (bottom). The encircled black cross 
represents the best--fit point. The inner and outer solid, black contours delimit the 
68\%\,C.L. and 95\%\,C.L. posterior regions respectively. We also illustrate the $5\sigma$ 
detection threshold of the LHC by the magenta dot-dashed curve for $\sqrt{s}=$14\,TeV 
and IL\,=\,1\,fb$^{-1}$, and by the solid blue curve for $\sqrt{s}=$14\,TeV and 
IL\,=\,100\,fb$^{-1}$.}
\label{fig:m0mhalf}
\end{figure*}
\end{center}
 %***********************************************************************************

%%%%%%%%%%%%%%%%%%%%%%%%%%%%%%%%%%%%%%%
\section{Results}
\label{sec:results}
%%%%%%%%%%%%%%%%%%%%%%%%%%%%%%%%%%%%%%%

%%%%%%%%%%%%%%%%%%%%%%%%%%%%%%%%%%%%%%%
\subsection{The cMSSM after null searches at the LHC}
\label{subsec:cMSSM}
%%%%%%%%%%%%%%%%%%%%%%%%%%%%%%%%%%%%%%%

We begin by displaying in Fig.\,\ref{fig:m0mhalf} the favoured region of the cMSSM 
parameter space (in terms of the Bayesian posterior pdf) in the ($m_0$, $m_{1/2}$) 
plane, where we have marginalized over all other parameters. The top and bottom 
rows show results for the flat and log priors respectively. From left to right, we have 
imposed current LHC constraints (left panels), and future constraints assuming that 
the LHC does not detect SUSY with $\sqrt{s}=$14\,TeV and IL\,=\,1\,fb$^{-1}$
(middle panels) and $\sqrt{s}=$14\,TeV and IL\,=\,100\,fb$^{-1}$  (right panels). This has been achieved 
by removing samples which lie below the $5\sigma$ detection region, depicted by 
the magenta dot-dashed curve for the LHC with $\sqrt{s}=$14\,TeV and IL\,=\,1\,fb$^{-1}$, 
and by the solid blue curve for $\sqrt{s}=$14\,TeV and IL\,=\,100\,fb$^{-1}$.

With the current LHC configuration, both the stau co-annihilation and $h$-pole regions 
remain inaccessible. The latter (visible as a small, vertical sliver of probability probability at 
small values of $m_{1/2}$) we see will become accessible with an integrated luminosity of 1\,fb$^{-1}$ 
and a centre of mass energy of 14\,TeV. For an LHC configuration of $\sqrt{s}=$14\,TeV 
and IL\,=\,100\,fb$^{-1}$, we see that the stau co--annihilation region displayed becomes 
completely accessible. In case of a lack of detection, the surviving posterior probability is completely 
confined to what, from here on, we call the funnel region, which encompasses the 
$A$-funnel and a small fraction of the focus point region which evades the current XENON-100 
constraints, since it is pushed to large $m_\chi$, for both choices of priors, as can be seen 
in the right-most panels of Fig.~\ref{fig:m0mhalf}. The percentage of the currently favoured 
cMSSM parameter space which would survive a lack of detection at the LHC is given in 
Table~\ref{tab:LHC_reach}, for both choices of priors. Those values can be interpreted 
as probabilities for the {\em nightmare} scenario in the cMSSM: for the LHC configuration 
with $\sqrt{s}=14$\,TeV and IL\,=\,100\,fb$^{-1}$, the probability of a non-detection lies between about 3\% 
and 36\%, depending on the choice of priors, with the flat prior giving a larger probability 
to the nightmare scenario (as a larger fraction of its posterior probability lies in the funnel 
region, beyond the reach of the LHC).

 %***********************************************************************************
\begin{table*}[t]
\centering
\begin{tabular}{|c|cc|}
\hline
Prior & \multicolumn{2}{c|}{LHC Configuration ($\sqrt{s}$, IL)}\\
 & 14\,TeV,\,1\,fb$^{-1}$ & 14\,TeV,\,100\,fb$^{-1}$\\
\hline
\hline
Flat  & 93.8 & 36.0 \\
Log  & 66.8 &  3.0 \\
\hline
\end{tabular}
\caption{Percentage of the currently favoured parameter space in the cMSSM that 
would survive a lack of SUSY detection at the LHC with the given configuration of 
centre of mass energy $\sqrt{s}$ and integrated luminosity IL. }
\label{tab:LHC_reach}
\end{table*}
 %***********************************************************************************
 %***********************************************************************************
\begin{center}
\begin{figure*}[t]
\includegraphics[width=0.32\textwidth,keepaspectratio,clip]{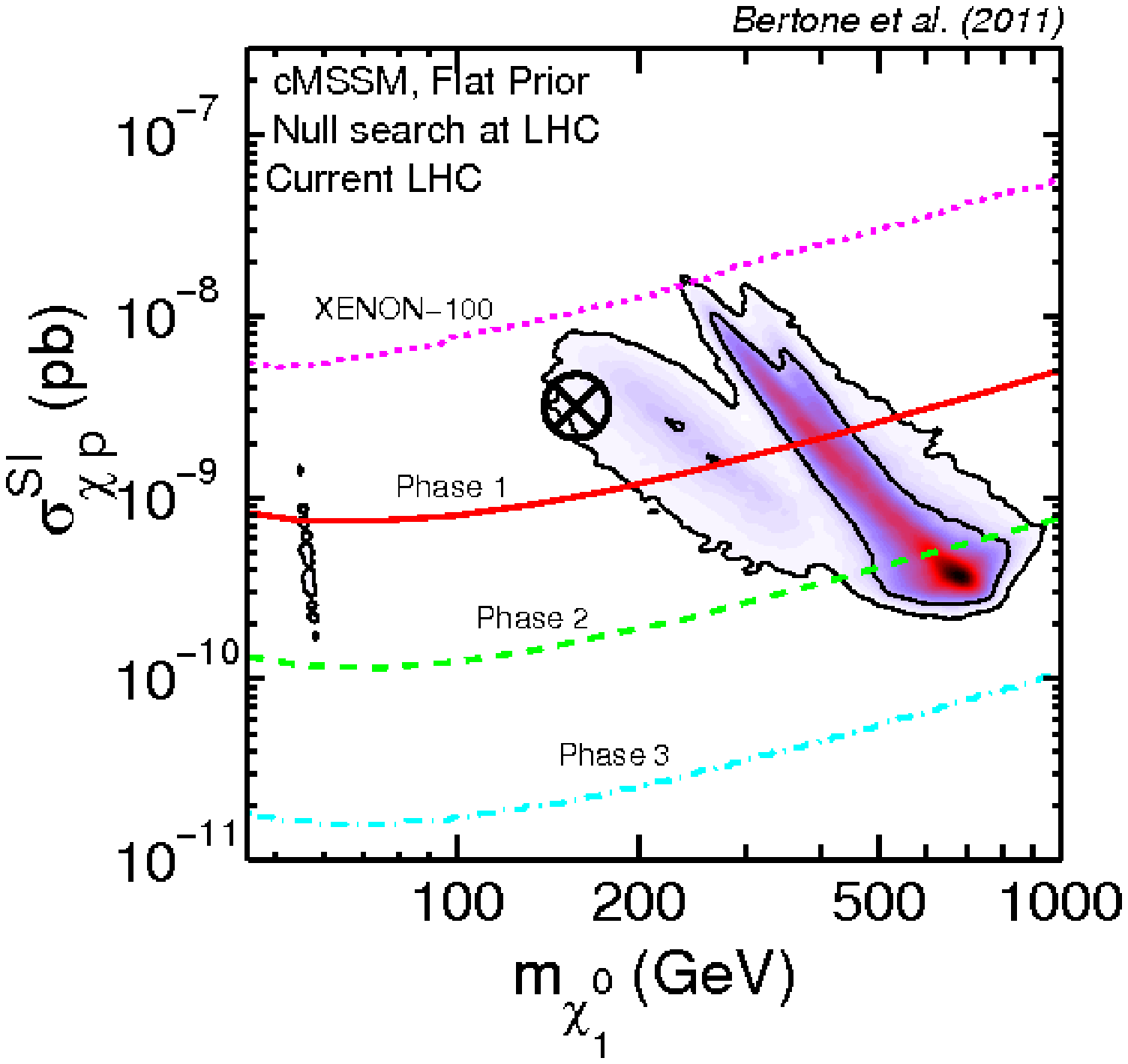}
\includegraphics[width=0.32\textwidth,keepaspectratio,clip]{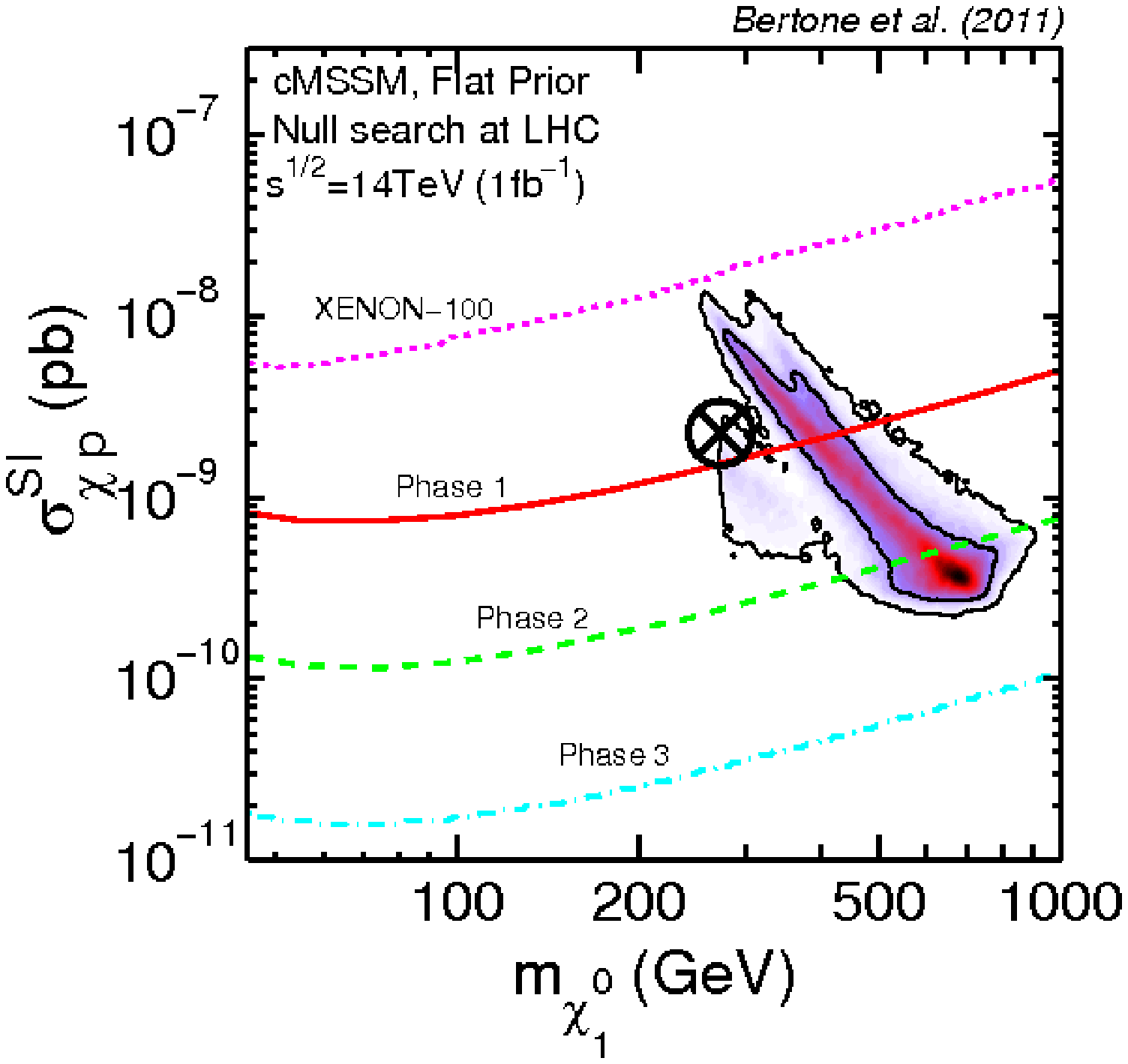}
\includegraphics[width=0.32\textwidth,keepaspectratio,clip]{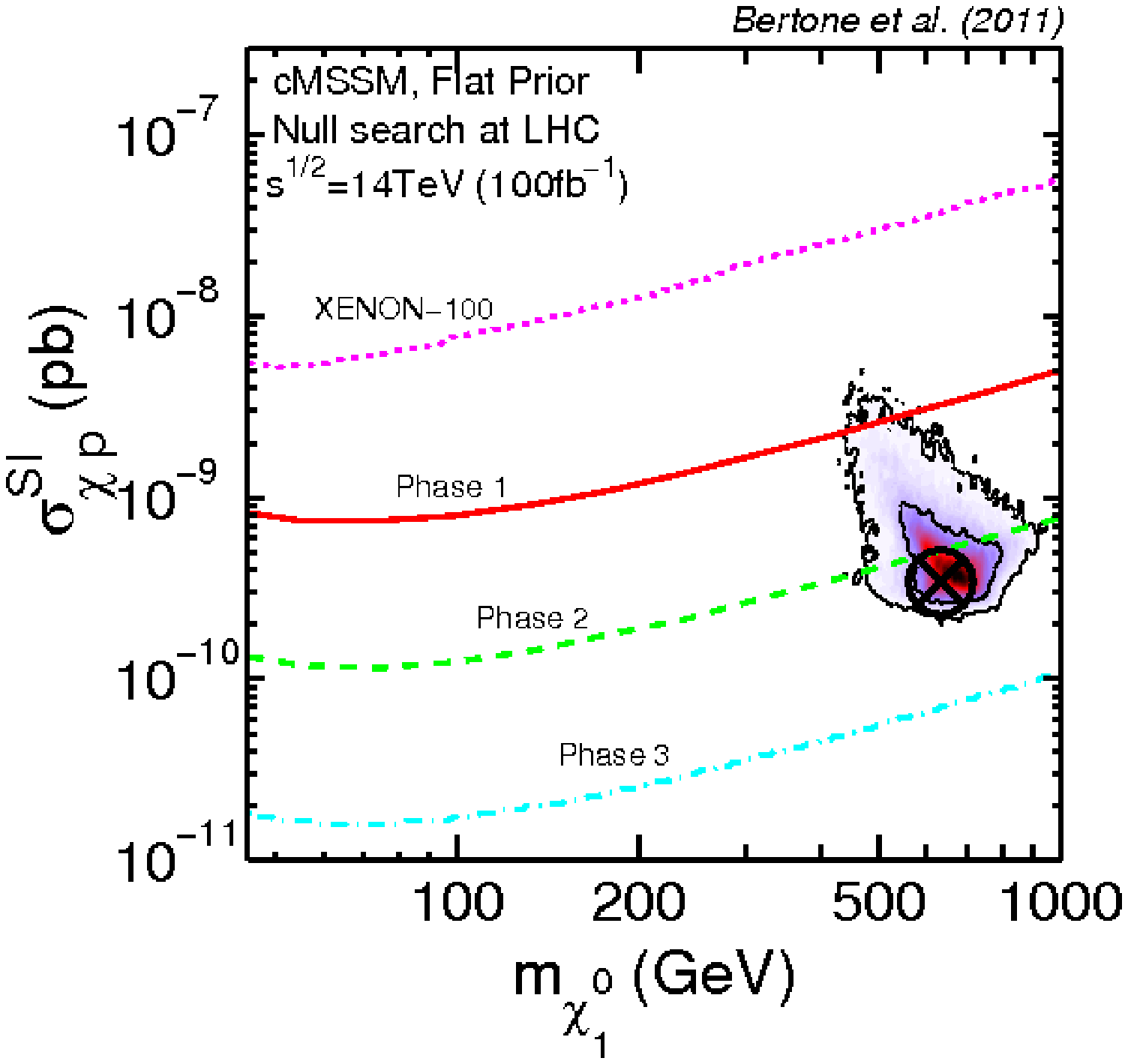}\\
\includegraphics[width=0.32\textwidth,keepaspectratio,clip]{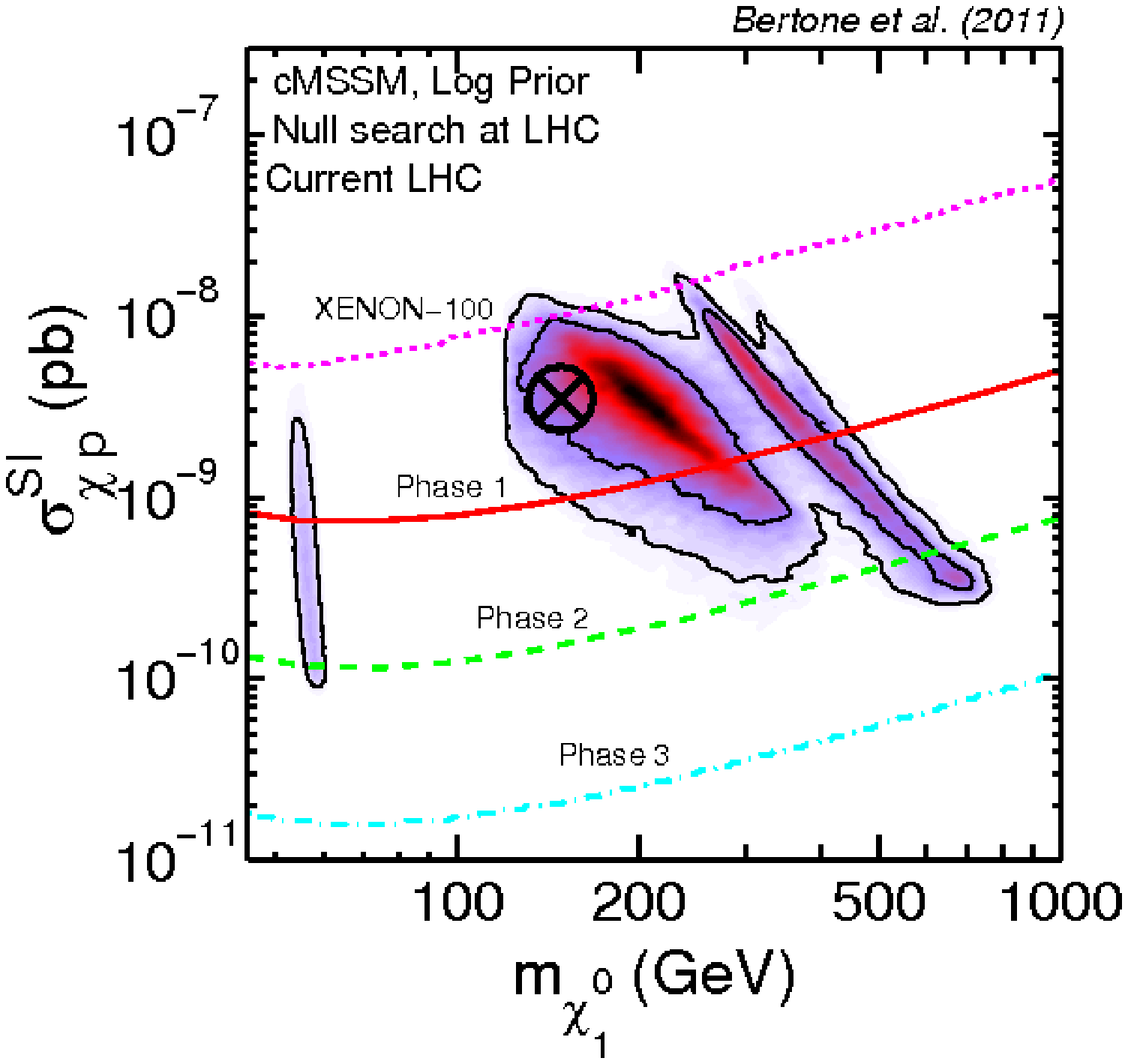}
\includegraphics[width=0.32\textwidth,keepaspectratio, clip]{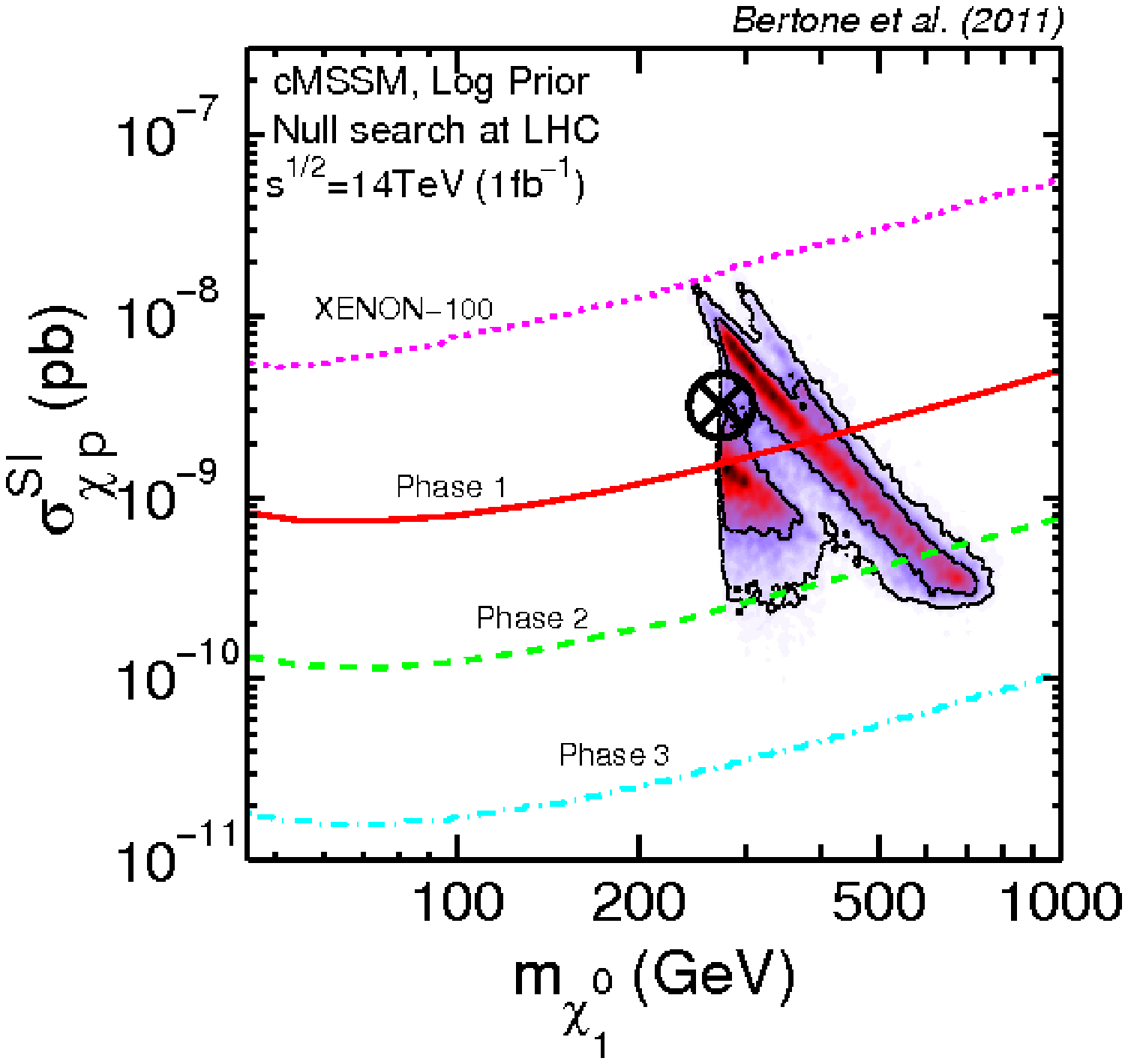}
\includegraphics[width=0.32\textwidth,keepaspectratio, clip]{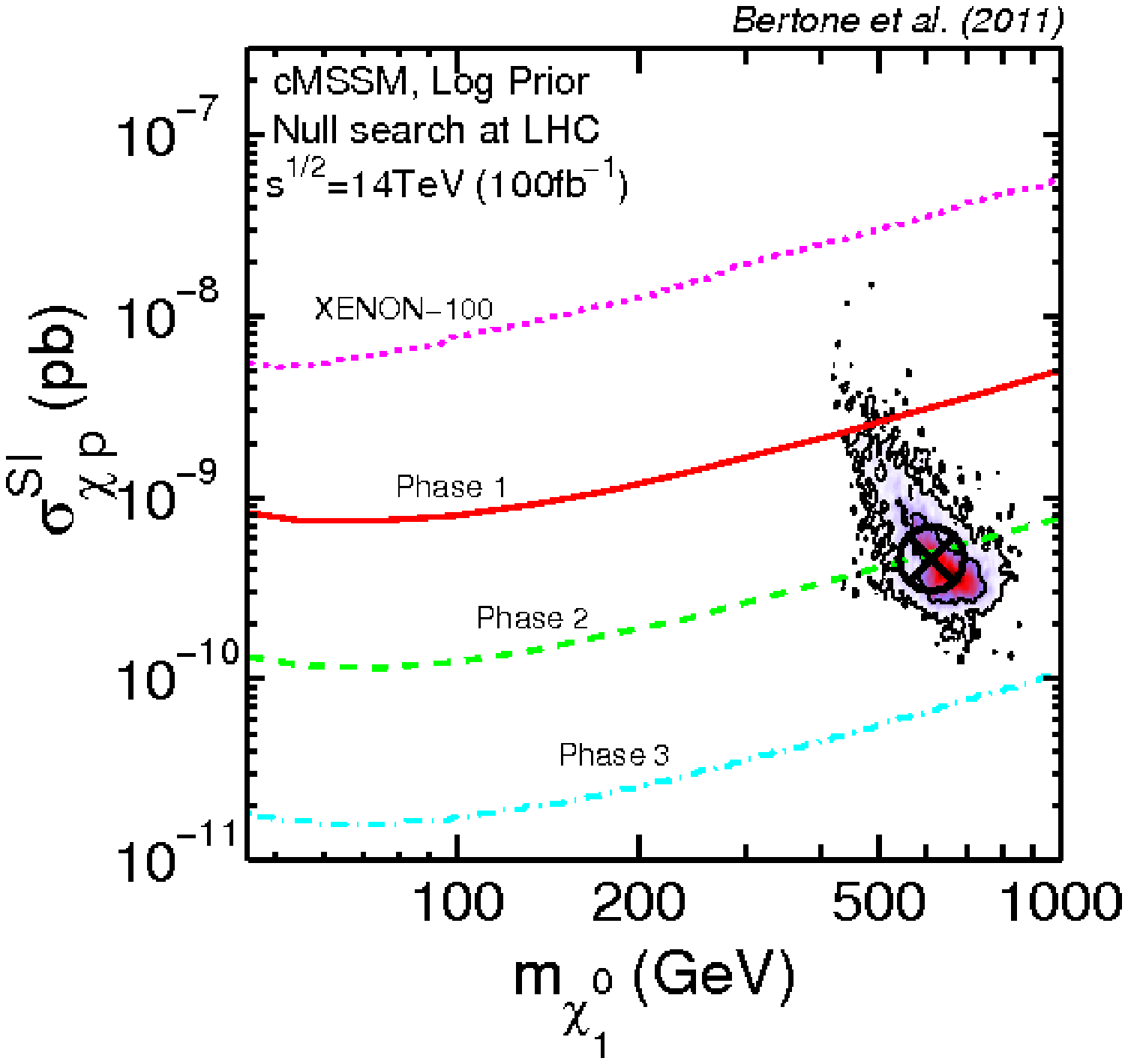}
\caption{Favoured region in the cMSSM once current constraints and future null 
searches at the LHC are taken into account, for flat priors (top) and log priors (bottom). 
We assume null searches at the LHC with the following combinations of $\sqrt{s}$ 
and integrated luminosities (from left to right): Current LHC, 14\,TeV and 1\,fb$^{-1}$, 
and 14\,TeV and 100\,fb$^{-1}$. The encircled black cross represents the best--fit 
point. The inner and outer solid, black contours delimit the 68\%\,C.L. and 95\%\,C.L. 
posterior regions respectively. We also show the current 90\% C.L. exclusion limit 
from XENON-100 (magenta dotted),  and the expected reach for for Phase 1 (solid 
red curve, expected to be reached by $\sim$2012), Phase 2 (dashed green curve) 
and Phase 3 (dash-dotted cyan curve, expected to to be reached around 2020) 
future direct detection experiments.}
\label{fig:mchisigsip_ansatz}
\end{figure*}
\end{center}
 %***********************************************************************************

%%%%%%%%%%%%%%%%%%%%%%%%%%%%%%%%%%%%%%%
\subsection{Implications for Direct detection}
\label{subsec:DirectDetection}
%%%%%%%%%%%%%%%%%%%%%%%%%%%%%%%%%%%%%%%

In this section, we investigate the impact of hypothetical null searches at the LHC 
for future ton-scale direct dark matter detectors. 

In order to establish the prospects for {\it directly} detecting cMSSM DM, we need 
to make some assumptions regarding the local neutralino density and velocity 
distribution. A consistent way to combine accelerator and DD data is to impose the 
`scaling Ansatz' discussed in \cite{Bertone:2010rv}, which rescales the local density 
of neutralinos according to their predicted cosmological relic density. Here, however, 
we focus on the regions of the cMSSM parameter space where the neutralino is the 
{\it sole} constituent of DM, by imposing a Gaussian likelihood on $\Omega_\chi h^2$ 
(see Table I), and therefore the scaling Ansatz becomes almost irrelevant.

For an easier comparison with the existing literature we have fixed the astrophysical 
parameters describing the density and velocity distribution of DM particles to the 
commonly adopted benchmark values: local CDM density 
$\rho_{\odot,{\rm CDM}}=0.4\,$GeV\,cm$^{-3}$; circular velocity $v_0=235$ km s$^{-1}$ 
and escape velocity $v_{esc}=550$ km s$^{-1}$ (see, e.g., \cite{Pato:2010zk} and 
references therein for a recent discussion of the astrophysical uncertainties relating 
to these quantities). We have also neglected hadronic uncertainties in WIMP-nucleon 
couplings, adopting for the light quarks contribution to the nucleon form factors the values $f_{Tu}=0.02698$, $f_{Td}=0.03906$ and $f_{Ts}=0.36$ \cite{Ellis:2008hf}. Accounting for such uncertainties by including them as 
nuisance parameters in the scan would only mildly change the numerical results 
presented here, but the main conclusions would remain unchanged, in agreement 
with the findings of \cite{Bertone:2010rv,Bertone:2011nj}.  
 %***********************************************************************************
\begin{table*}[t]
\centering
\begin{tabular}{|c|c|ccc|}
\hline
Experiment & Prior & \multicolumn{3}{c|}{LHC Configuration ($\sqrt{s}$, IL)}\\
& & Current LHC & 14\,TeV,\,1\,fb$^{-1}$ & 14\,TeV,\,100\,fb$^{-1}$\\
\hline
\hline
Direct detection, Phase 1 & &30.2 &  23.5 &  2.5 \\
Direct detection, Phase 2 & Flat &73.4 & 69.7 & 35.5 \\
Direct detection, Phase 3 & &100.0 & 100.0 & 100.0 \\
IceCube plus DeepCore & & 9.6 &  6.9 &  0.5 \\
\hline
\hline
Direct detection, Phase 1 & &68.6 & 35.7 &  1.2 \\
Direct detection, Phase 2 & Log &95.8 & 89.0 & 31.1 \\
Direct detection, Phase 3 & &100.0 & 100.0 & 100.0 \\
IceCube plus DeepCore & &16.8 & 11.1 &  0.2 \\
\hline
\end{tabular}
\caption{Fraction of the cMSSM parameter space surviving null searches at the 
LHC which can be probed by direct detection experiments or by IceCube (the 
latter after 5 years of observation). Those values represent the probability of 
discovery by the corresponding experiment (for each prior choice) assuming the 
LHC fails to discover SUSY at the given energy and integrated luminosity.}
\label{tab:survival}
\end{table*}
 %***********************************************************************************

In Fig.\,\ref{fig:mchisigsip_ansatz}, we display the results of the posterior probability 
distribution of the cMSSM in the ($m_{\chi}$, $\sigsip$) plane following our scan 
corresponding to null searches at the LHC with the same priors and combinations 
of $\sqrt{s}$ and integrated luminosities as displayed in Fig.\,\ref{fig:m0mhalf}. For 
comparison, the expected sensitivity of upcoming direct detection experiments is 
also shown for the following experimental setups:
 %***********************************************************************************
\begin{itemize}
\item {\bf Phase 1:} experiments will probe cross sections down to $\sigsip\sim10^{-9}$\,pb, 
corresponding to the projected sensitivity of SuperCDMS at SNOLab \cite{SuperCDMS} 
with a detector target mass of 27\,kg operating for 1 year (solid red curve), roughly 
equivalent to the reach of XENON-100 by the end of 2012. 
\item {\bf Phase 2:} experiments will probe cross sections down to $\sigsip\sim10^{-10}$\,pb, 
corresponding to the projected sensitivity of SuperCDMS at SNOLab with a detector 
target mass of 145\,kg  running for 3 years (dashed green curve).
\item {\bf Phase 3:} experiments will probe cross sections down to $\sigsip\sim10^{-11}$\,pb, 
corresponding to the projected sensitivity of SuperCDMS at SNOLab with a detector 
target mass of 1500\,kg running for 4 years (dashed green curve), that should become 
available by 2021 \cite{SuperCDMS}. The Xenon1T is expected to reach a sensitivity 
of $\sigsip = 4 \times 10^{-11}$\,pb by 2015 \cite{Xenon1T}, and therefore to go down 
to $\sigsip \sim 10^{-11}$pb on a similar timescale as SuperCDMS.
\end{itemize}
 %***********************************************************************************
 
The left panels of Fig.~\ref{fig:mchisigsip_ansatz} show the favoured cMSSM region 
with current LHC data, for flat (top) and log (bottom) priors. Table\,\ref{tab:survival} 
summarizes the percentage of our points that reside outside of the reach of the LHC, 
i.e., in the nightmare scenario (for a given configuration) that are detectable for each 
of the experimental direct detection phases described above.  

The $h$-pole, stau co--annihilation and funnel regions displayed in the corresponding 
plots of Fig.\,\ref{fig:m0mhalf} can be observed in Fig.\,\ref{fig:mchisigsip_ansatz} as 
``islands" in the parameter space bound by 68\%\,C.L. contours and spanning the 
approximate mass ranges: $50\,{\rm GeV}\lesssim\mchi\lesssim60\,{\rm GeV}$, 
$100\,{\rm GeV}\lesssim\mchi\lesssim500\,{\rm GeV}$ and $200\,{\rm GeV}\lesssim\mchi\lesssim1\,{\rm TeV}$ 
respectively, with the best-fit point from current data (left panels) being located in 
the stau co--annihilation region for both priors.

Large portions of these three regions are within reach of Phase 1 direct detection 
experiments, which will be able to detect between 30\% and 68\% (depending on 
the choice of priors) of the parameter space currently outside the reach of the LHC. 
For both choices of priors, we observe that Phase 2 experiments will cover a 
substantially larger fraction of the currently surviving parameter space, between 73\% 
and 95\% for flat and log priors, respectively. Phase 3 experiments are expected to 
cover entirely the most probable region of the cMSSM currently inaccessible to the 
LHC, a result that, as we shall see, holds true also for the ultimate reach of the LHC. 

After its $\sim$18 month shutdown period starting in 2013, the LHC will be brought 
up to $\sqrt{s}=14\,{\rm TeV}$. The central panel of Fig.\,\ref{fig:mchisigsip_ansatz} 
shows the implications for direct detection assuming IL\,=\,1\,fb$^{-1}$. The LHC data 
in this case will rule out the $h$-pole region and start to cut in to the stau co-annihilation 
region, pushing the neutralino mass to larger values (see Fig.\,\ref{fig:mchi} below) and 
substantially reducing the percentage of surviving points within the reach of Phase 1 
DD experiments, from 30\% to 23\% for flat priors, and from 68\% to 35\% for log priors. 
The prospects for detection with Phase 2 DD experiments remain fairly stable, while, 
again, Phase 3 DD experiments can probe all of the surviving parameter space.
 %***********************************************************************************
\begin{center}
\begin{figure*}[t]
\includegraphics[width=0.32\textwidth,keepaspectratio, clip]{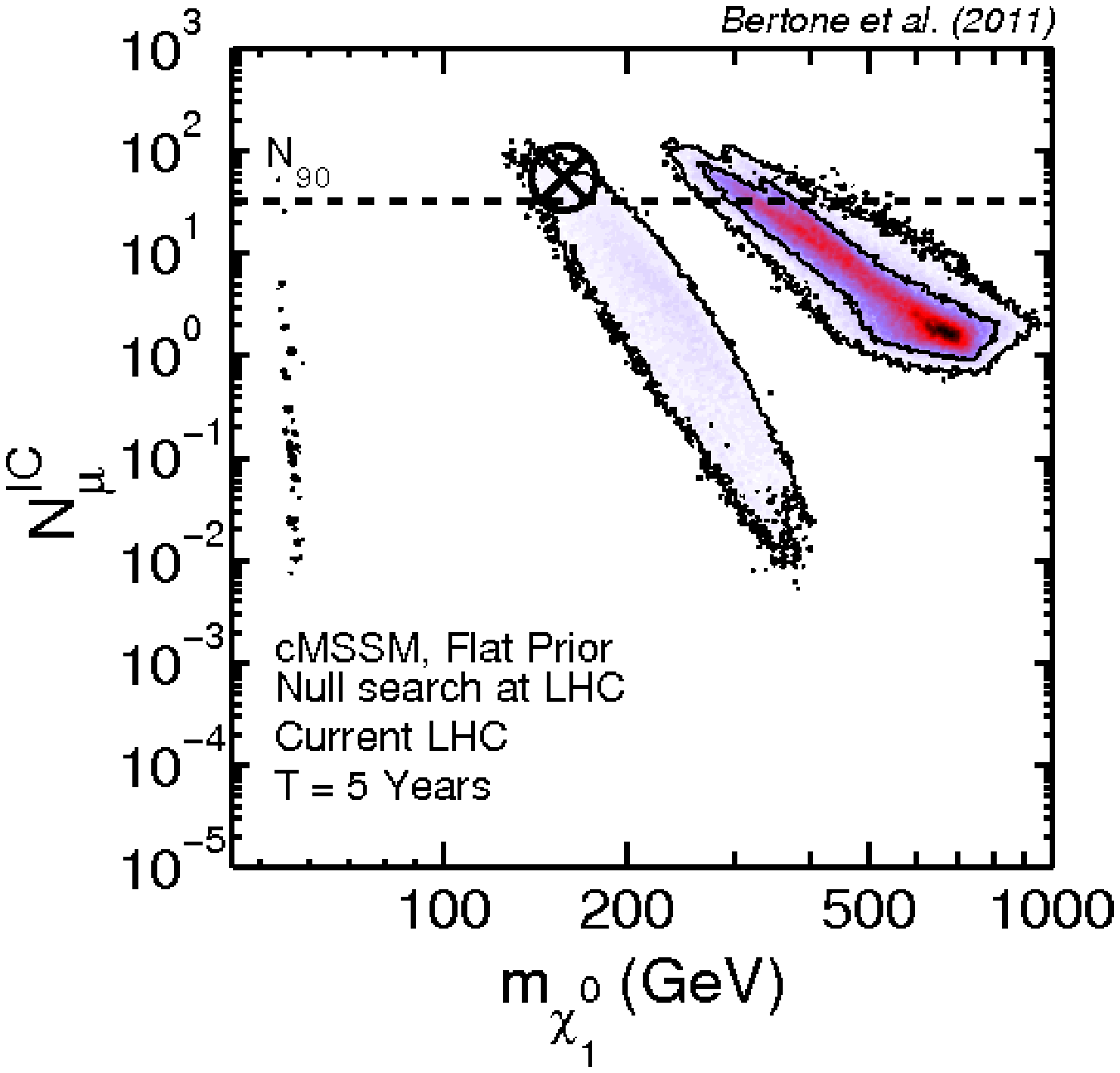}
\includegraphics[width=0.32\textwidth,keepaspectratio, clip]{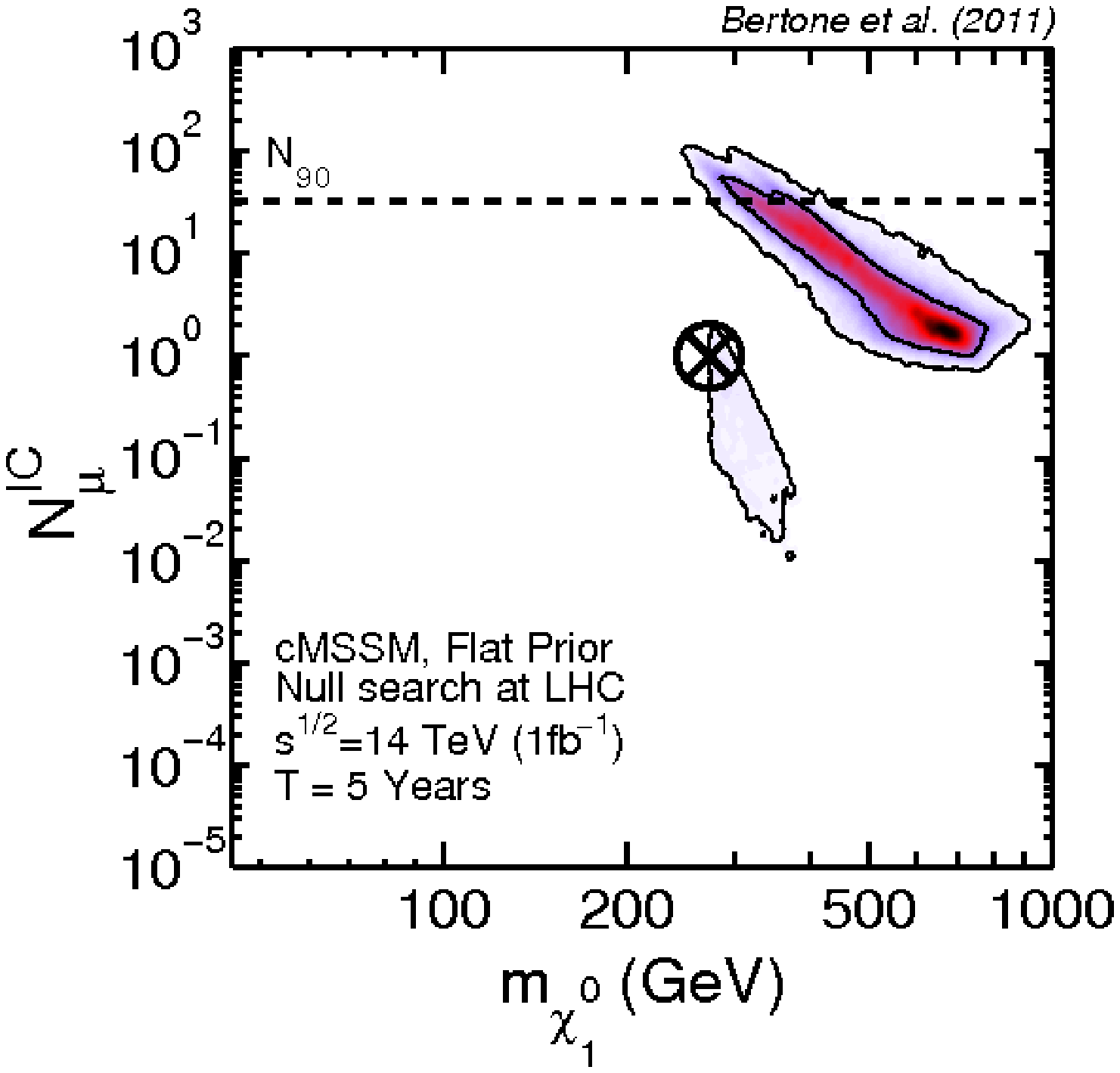}
\includegraphics[width=0.32\textwidth,keepaspectratio, clip]{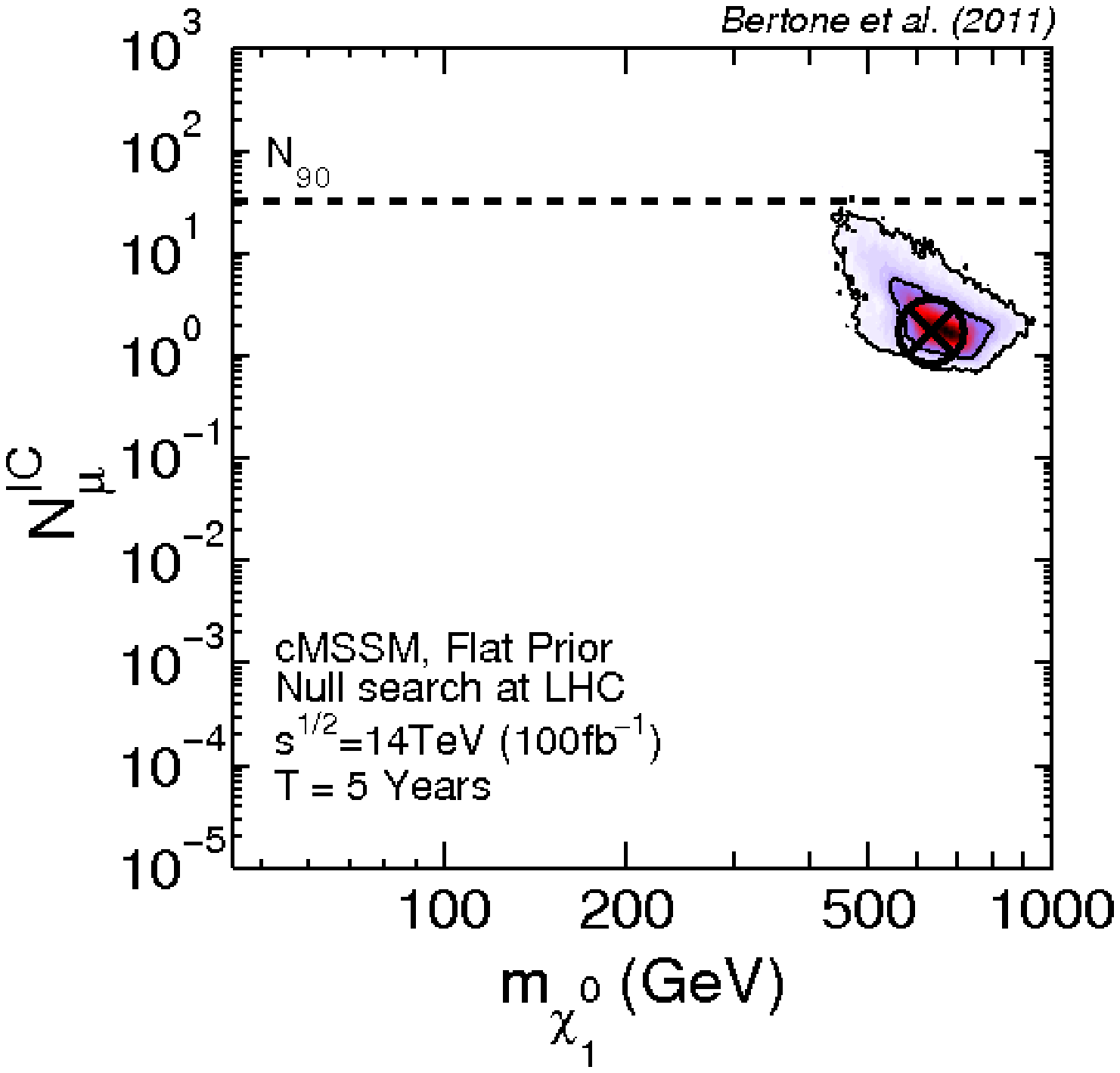}\\
\includegraphics[width=0.32\textwidth,keepaspectratio, clip]{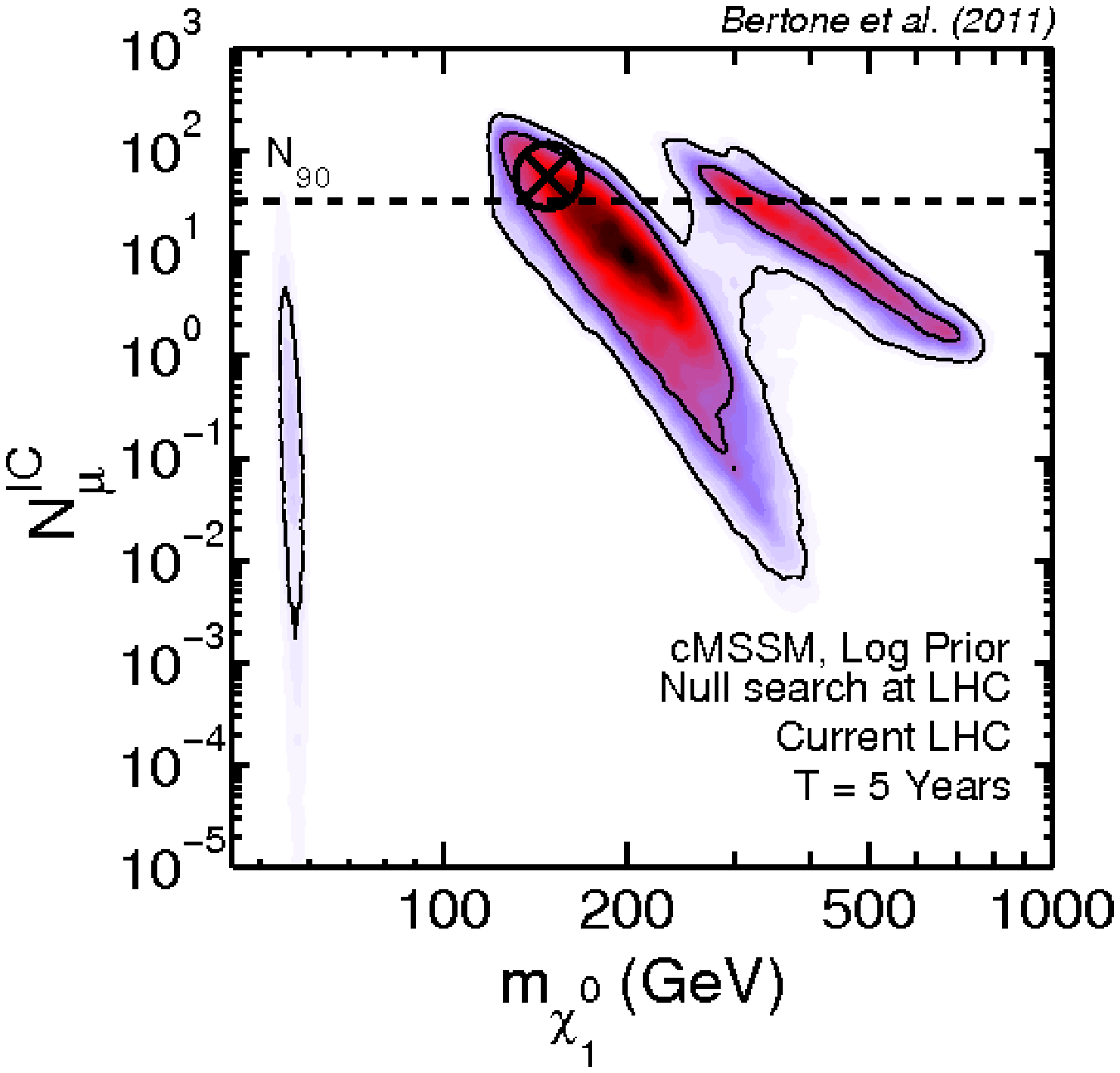}
\includegraphics[width=0.32\textwidth,keepaspectratio, clip]{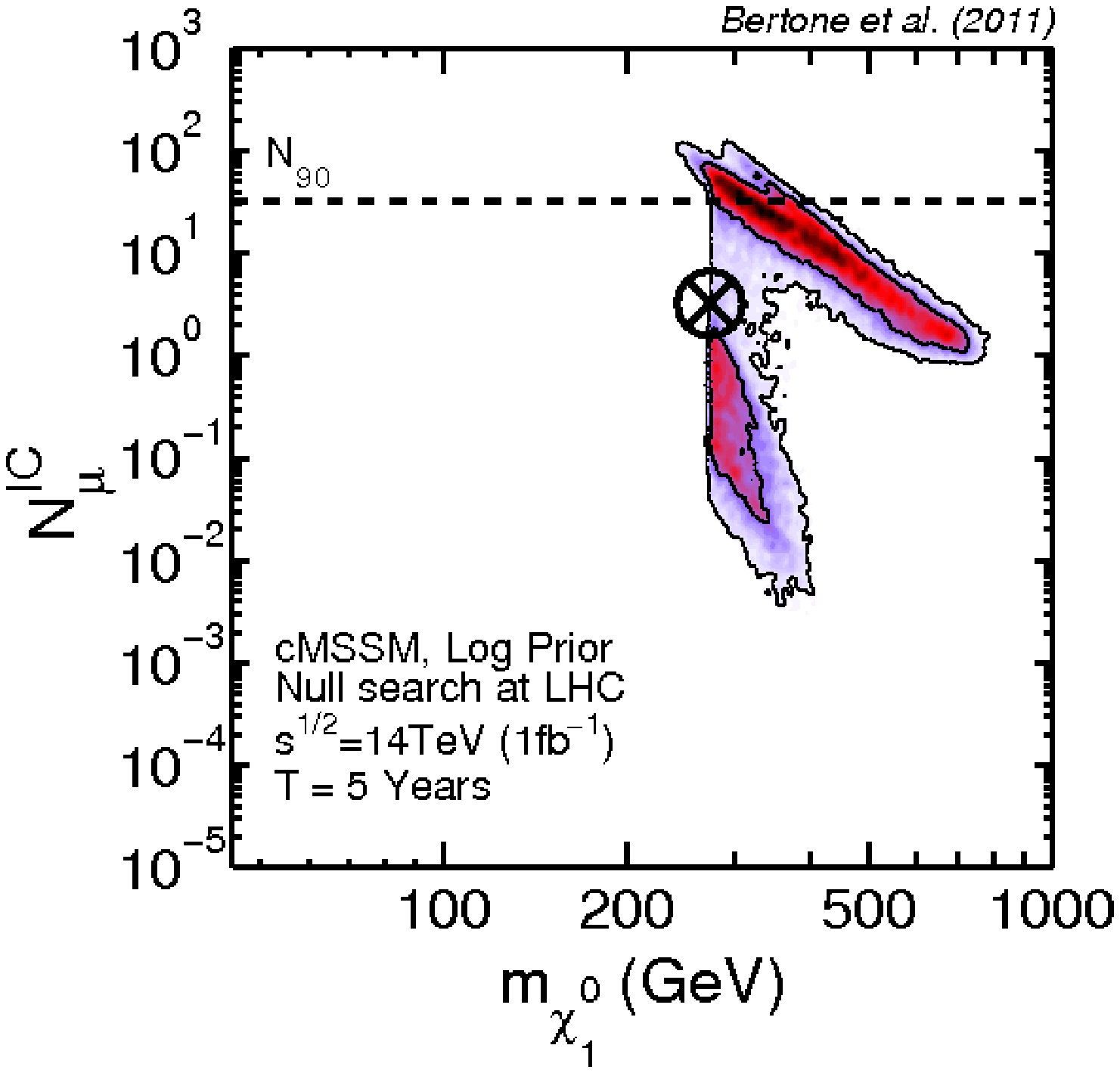}
\includegraphics[width=0.32\textwidth,keepaspectratio, clip]{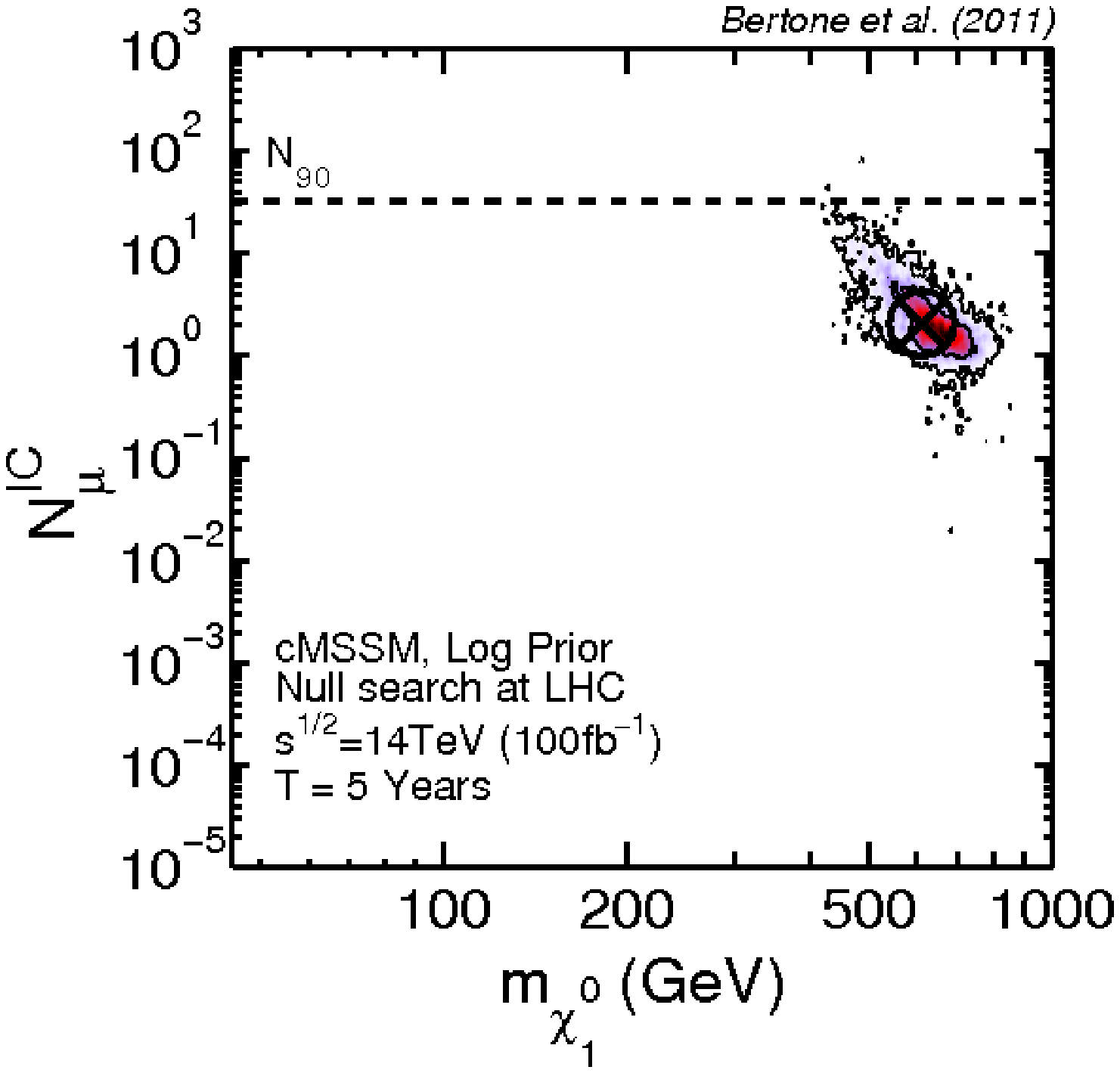}\\
\caption{Posterior probability distribution of the cMSSM, displayed in the 
($\mchi$, N$_{\mu}^{\rm IC}$) plane,  corresponding to null searches by the 
LHC with combinations of $\sqrt{s}$ and integrated luminosities (from left to right): 
Current LHC, 7\,TeV and 1\,fb$^{-1}$, 14\,TeV and 1\,fb$^{-1}$, and 14\,TeV and 
100\,fb$^{-1}$, when using flat priors (top) and log priors (bottom). The number 
of muon events $N_{\mu}^{\rm IC}$ assumes 5\,years of observations by IceCube 
plus DeepCore. The encircled black cross represents the best--fit point. The inner 
and outer solid, black contours delimit the 68\%\,C.L. and 95\%\,C.L. posterior 
regions respectively. The horizontal dashed line gives the $90\%\,$C.L. detection 
threshold, and the encircled black cross is the best--fit in each scenario.}
\label{fig:spec_g}
\label{fig:mchimuon}
\end{figure*}
\end{center}
 %***********************************************************************************

Lastly, for $\sqrt{s}=$14\,TeV and IL\,=\,100\,fb$^{-1}$, as expected from Fig.\,\ref{fig:m0mhalf}, 
for both flat and log prior scans the $h$-pole and stau co--annihilation regions 
are now fully accessible to the LHC, with the inaccessible part of the funnel region 
being quite similar for either set of priors. In both cases, we can see from Table\,\ref{tab:survival} 
that, whilst a small percentage (of order $1-2\%$) of points are accessible to the 
Phase 1 experiments. It is clear that Phase 2 detectors will be necessary in order 
to have a significant chance (i.e., of order $\sim30\%$, independently of the choice 
of priors) of discovering dark matter in the context of the cMSSM, whose gaugino and scalar mass parameters would in this case be located in the high mass funnel region, 
if it evades discovery by the LHC in the long term. 

Interestingly, we find that even in the case of null searches at the LHC with 
$\sqrt{s}=$14\,TeV and IL\,=\,100\,fb$^{-1}$, i.e., after many years of data taking 
after 2014, 100\% of the cMSSM posterior will be probed by Phase 3 DD experiments, 
that will become available on a similar timescale. 

 %***********************************************************************************
\begin{center}
\begin{figure*}[t]
\includegraphics[width=0.32\textwidth,keepaspectratio, clip]{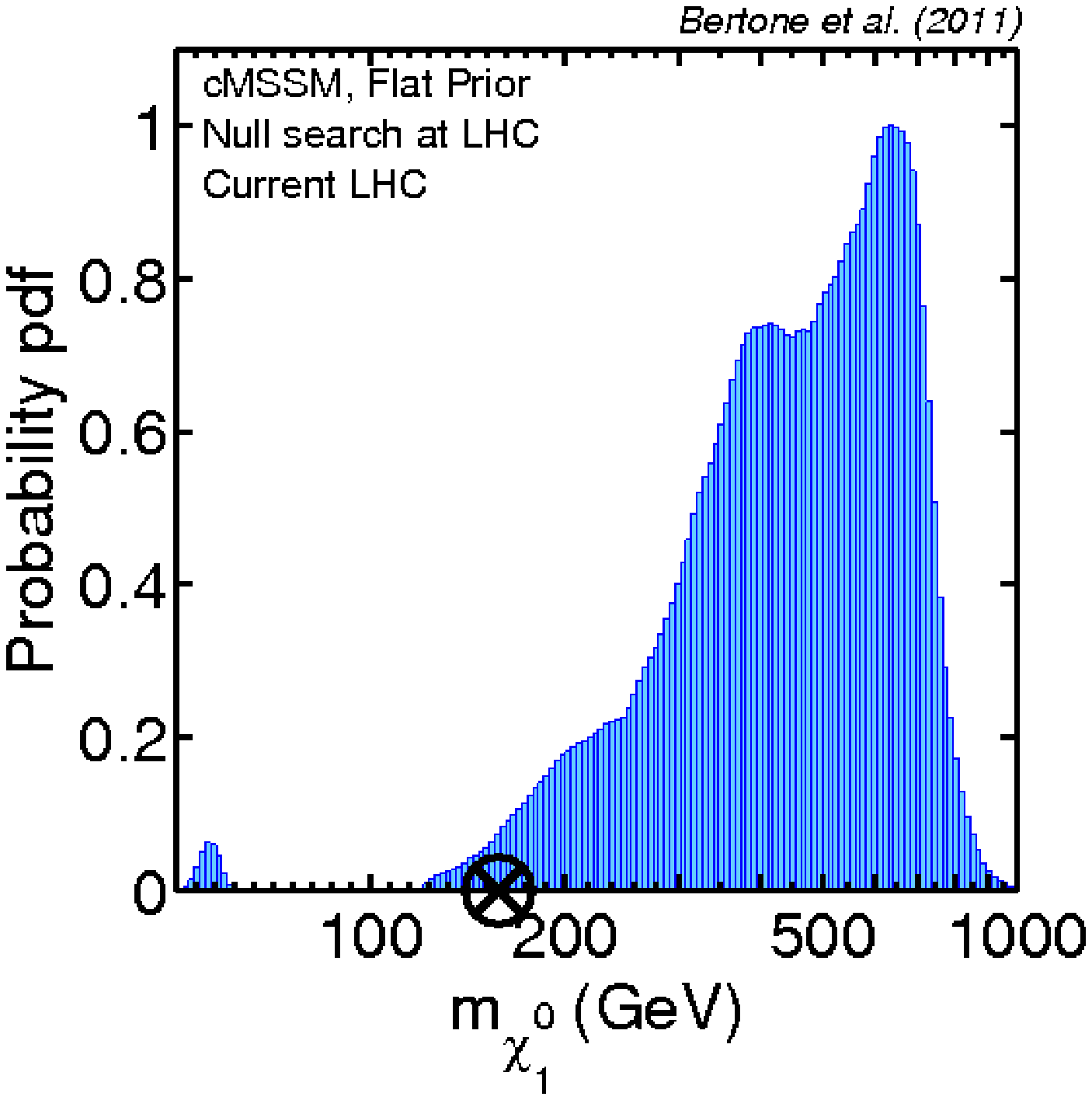}
\includegraphics[width=0.32\textwidth,keepaspectratio, clip]{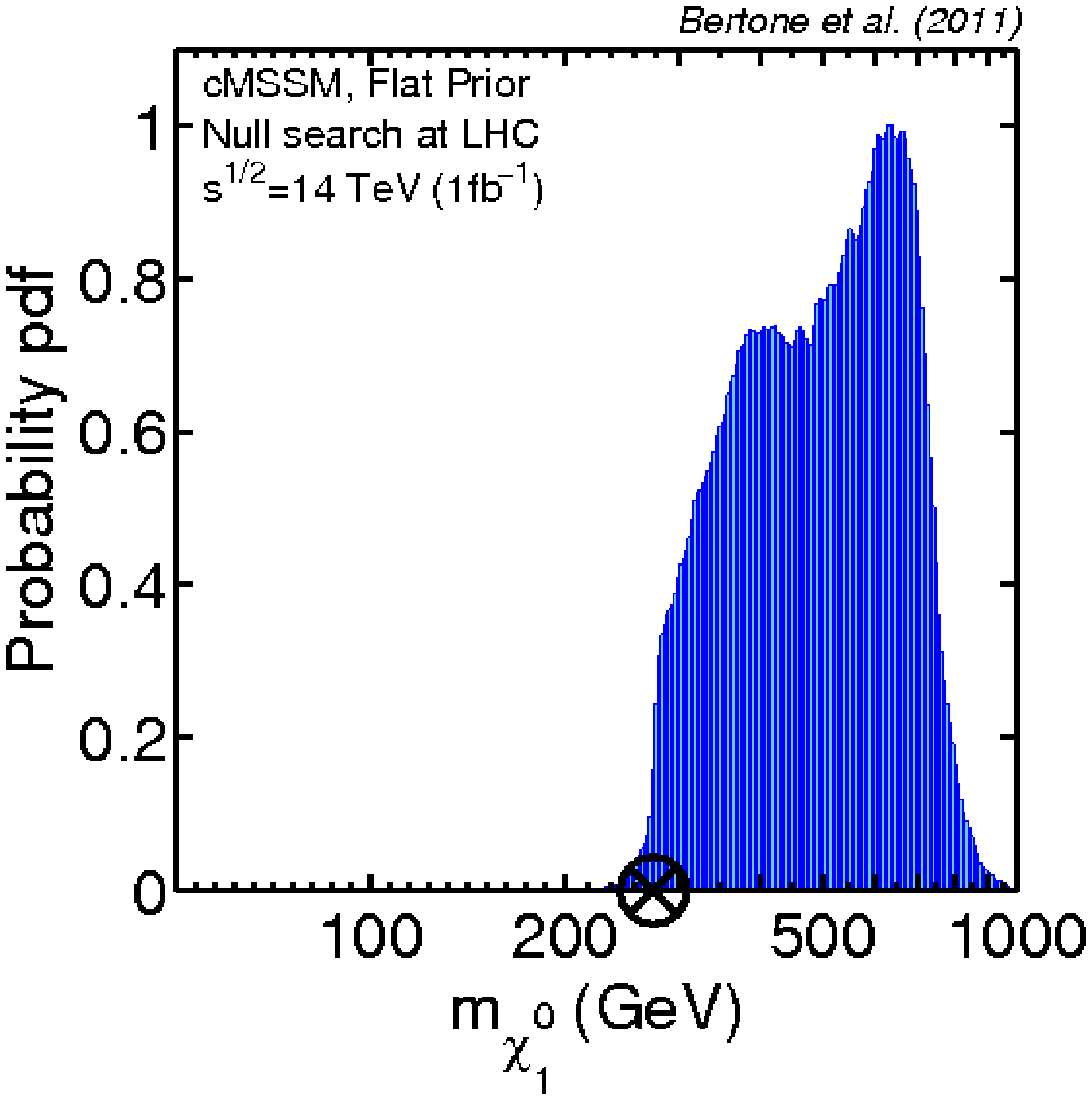}
\includegraphics[width=0.32\textwidth,keepaspectratio, clip]{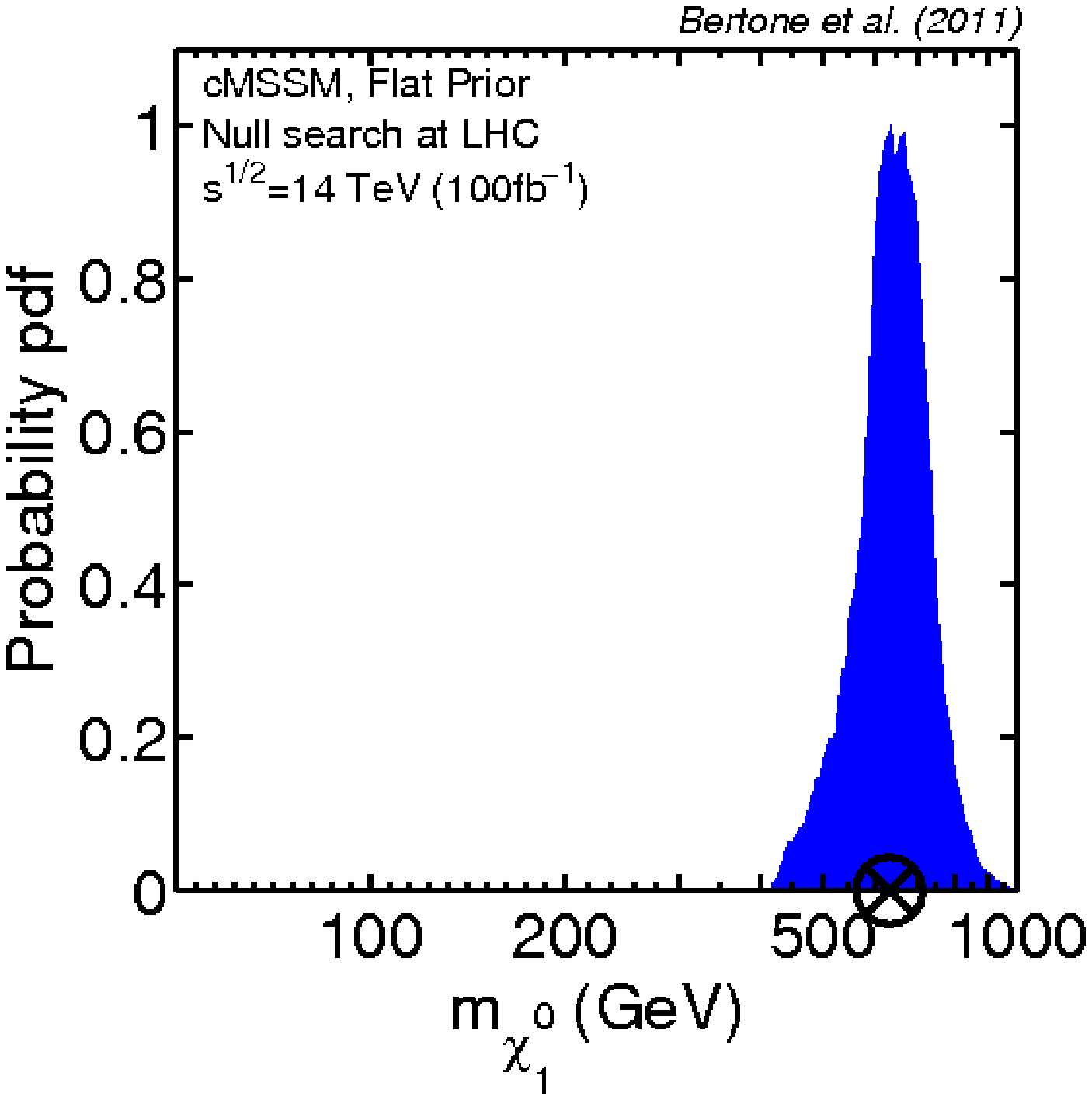}\\
\includegraphics[width=0.32\textwidth,keepaspectratio, clip]{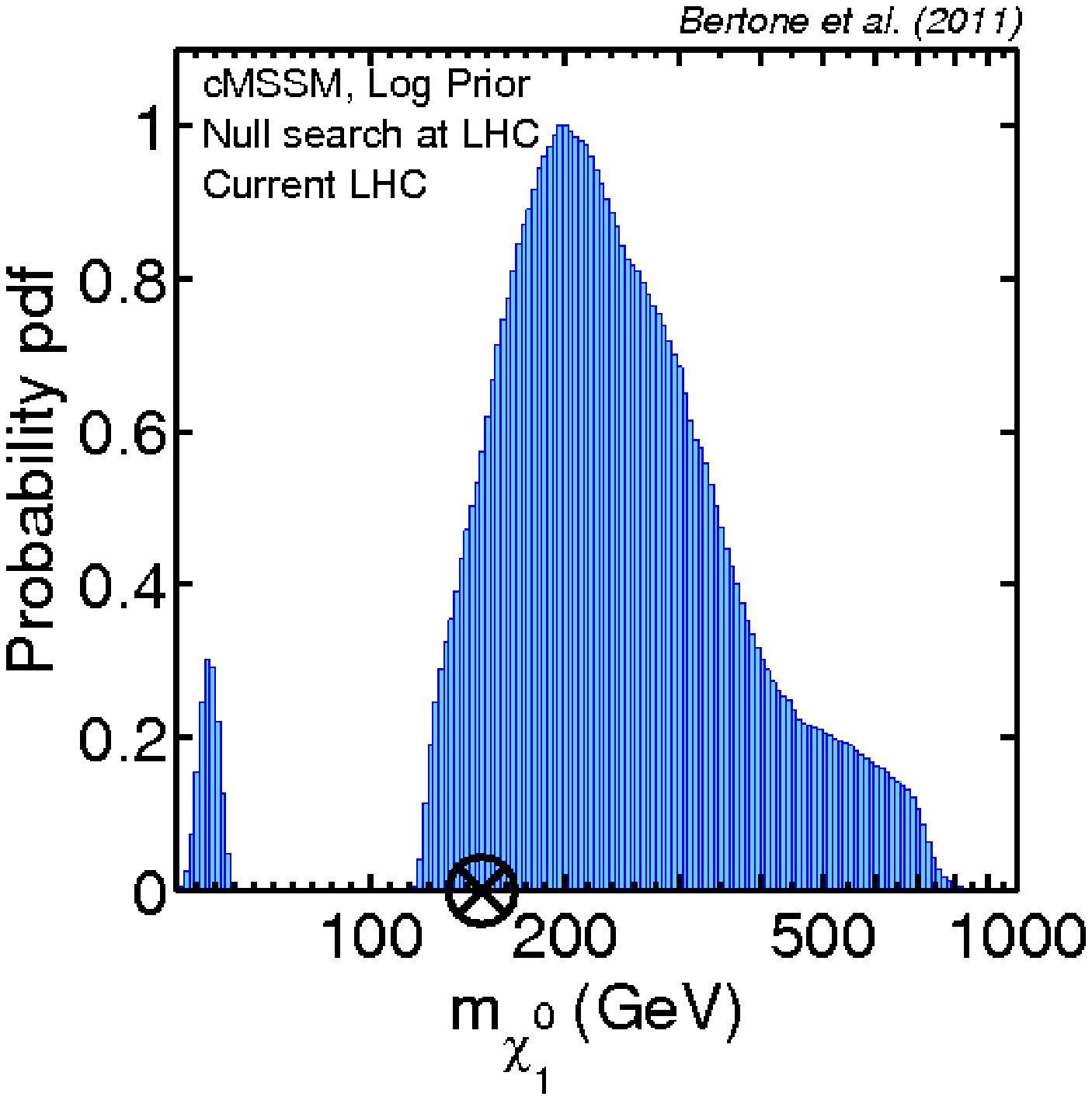}
\includegraphics[width=0.32\textwidth,keepaspectratio, clip]{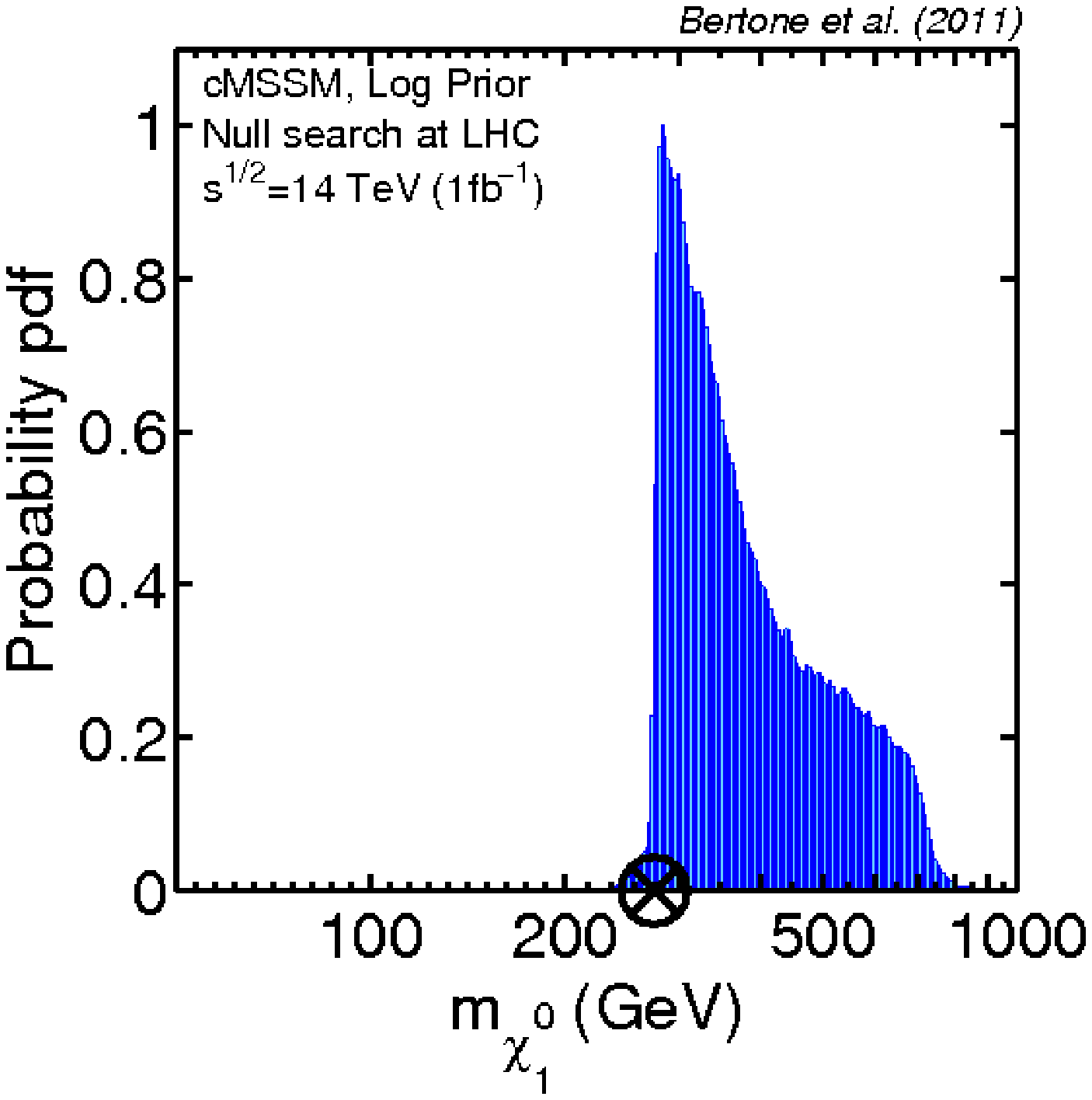}
\includegraphics[width=0.32\textwidth,keepaspectratio, clip]{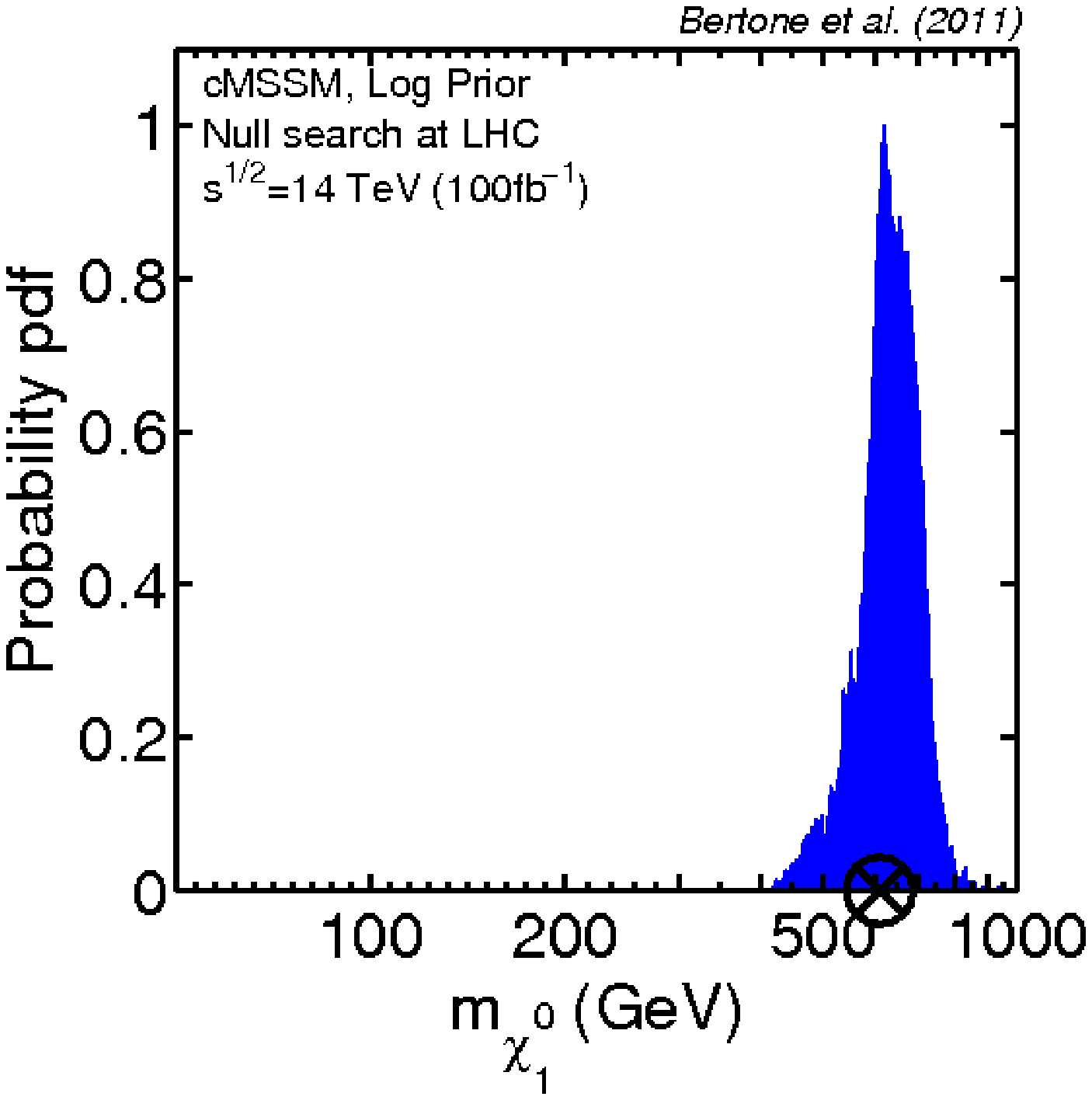}\\
\caption{1D pdf for the neutralino mass $\mchi$ for flat priors (top) and log priors 
(bottom), assuming Current LHC constraints (left), null searches at the LHC with 
$\sqrt{s}=$14\,TeV and IL\,=\,1\,fb$^{-1}$ (middle) and null searches at the LHC 
with $\sqrt{s}=$14\,TeV and IL\,=\,100\,fb$^{-1}$ (right panels). The encircled black 
cross represents the best--fit point. \label{fig:mchi}}
\end{figure*}
\end{center}
 %***********************************************************************************

%%%%%%%%%%%%%%%%%%%%%%%%%%%%%%%%%%%%%%%
\subsection{Implications for indirect detection with the IceCube neutrino telescope}
\label{subsec:IndirectDetection}
%%%%%%%%%%%%%%%%%%%%%%%%%%%%%%%%%%%%%%%

In this section we examine the prospects for detecting DM in the cMSSM with 
indirect searches in the nightmare scenario, focusing on the detection of high energy 
neutrinos from the Sun. Dark Matter particles are expected to scatter off nuclei in 
the Sun, and to sink at its centre, where they can subsequently annihilate. Among 
the annihilation products only neutrinos can escape, and produce a neutrino flux 
on Earth that is currently searched for with the IceCube neutrino telescope, located 
at the South Pole (see, e.g., \cite{Halzen:2009vu}). IceCube is now fully built, with 
80 strings in total, and it has been recently supplemented by a more densely 
instrumented region (consisting of six additional strings) at its core, called DeepCore 
(see, e.g., \cite{Ahrens:2003ix, Wiebusch:2009jf}), which has the effect of lowering 
its neutrino energy threshold, $E^{\rm th}_{\nu}$, to approximately 10\,{\rm GeV} and 
increasing the effective area for low-energy events~\cite{cowen:2009}.

The number of neutrino--induced muon events, $N_{\mu}^{\rm IC}$, observable by 
IceCube from the direction of the Sun can be computed as 
 %***********************************************************************************
\begin{equation}
N_{\mu}^{\rm IC}=\int_{\Delta\Omega_{\odot}}{\rm d}\Omega\int^\infty_{E_{\nu}^{\rm th}}
A_{\nu}^{\rm eff}(E_{\nu})\frac{{\rm d}\Phi_{\nu}}{{\rm d}E_{\nu}}{\rm d}E_{\nu}
\label{eq:Nmu}
\end{equation}
 %***********************************************************************************
where  $\frac{{\rm d}\Phi_{\nu}}{{\rm d}E_{\nu}}$ is the incident muon neutrino flux 
at the detector, ${E_{\nu}^{\rm th}}$ is the energy threshold, and $A_{\nu}^{\rm eff}(E_{\nu})$ 
is the muon neutrino effective area of the detector for muon neutrinos with energy 
$E_{\nu}$, averaged over the northern hemisphere. The flux is then integrated 
over the solid angle $\Delta\Omega_{\odot}$ subtended by the Sun, corresponding 
to a cone with half-angle of approximately $0.3^{\circ}$.
 %***********************************************************************************
\begin{center}
\begin{figure*}[t]
\includegraphics[width=0.485\textwidth,keepaspectratio, clip]{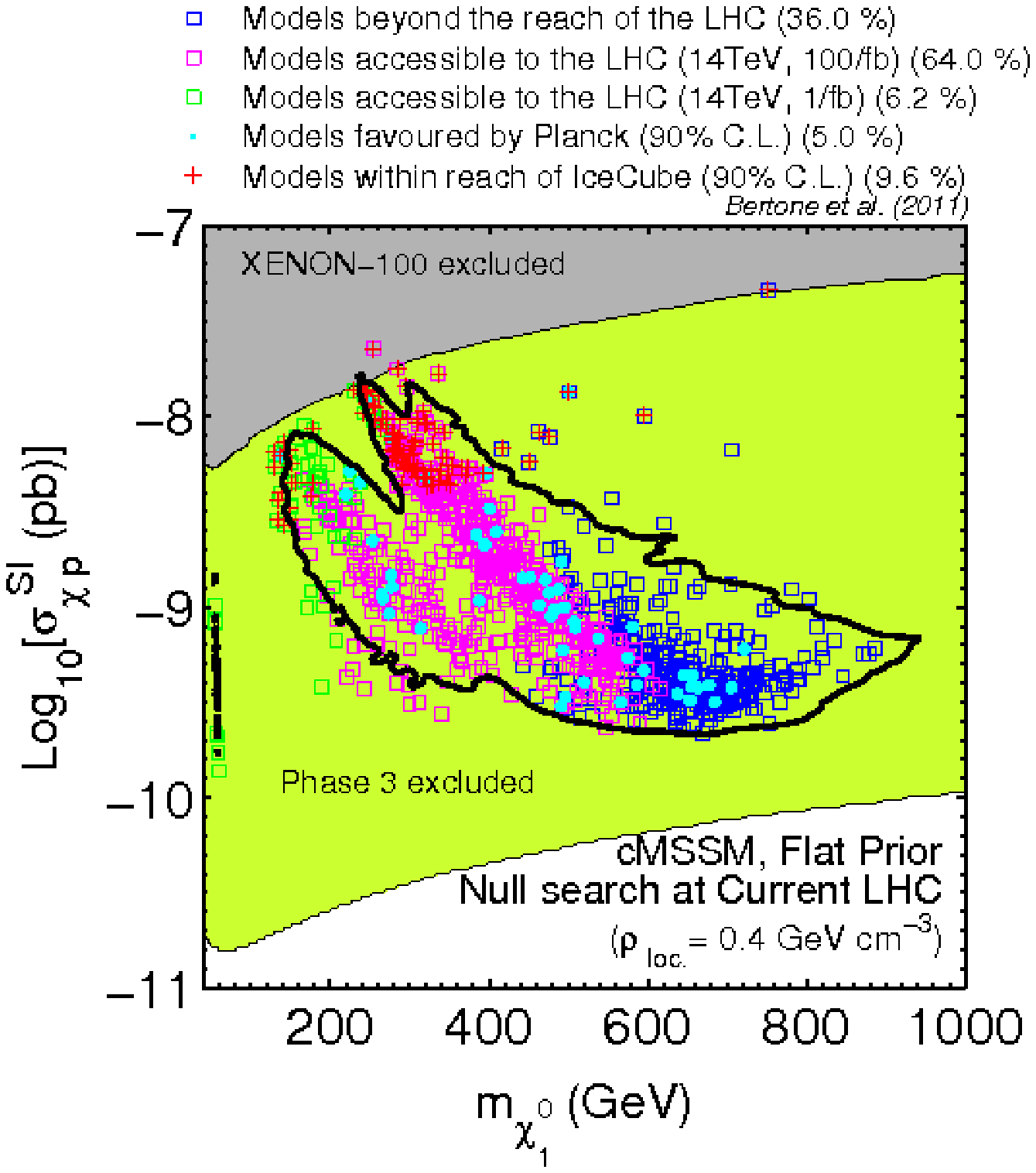}
\includegraphics[width=0.495\textwidth,keepaspectratio, clip]{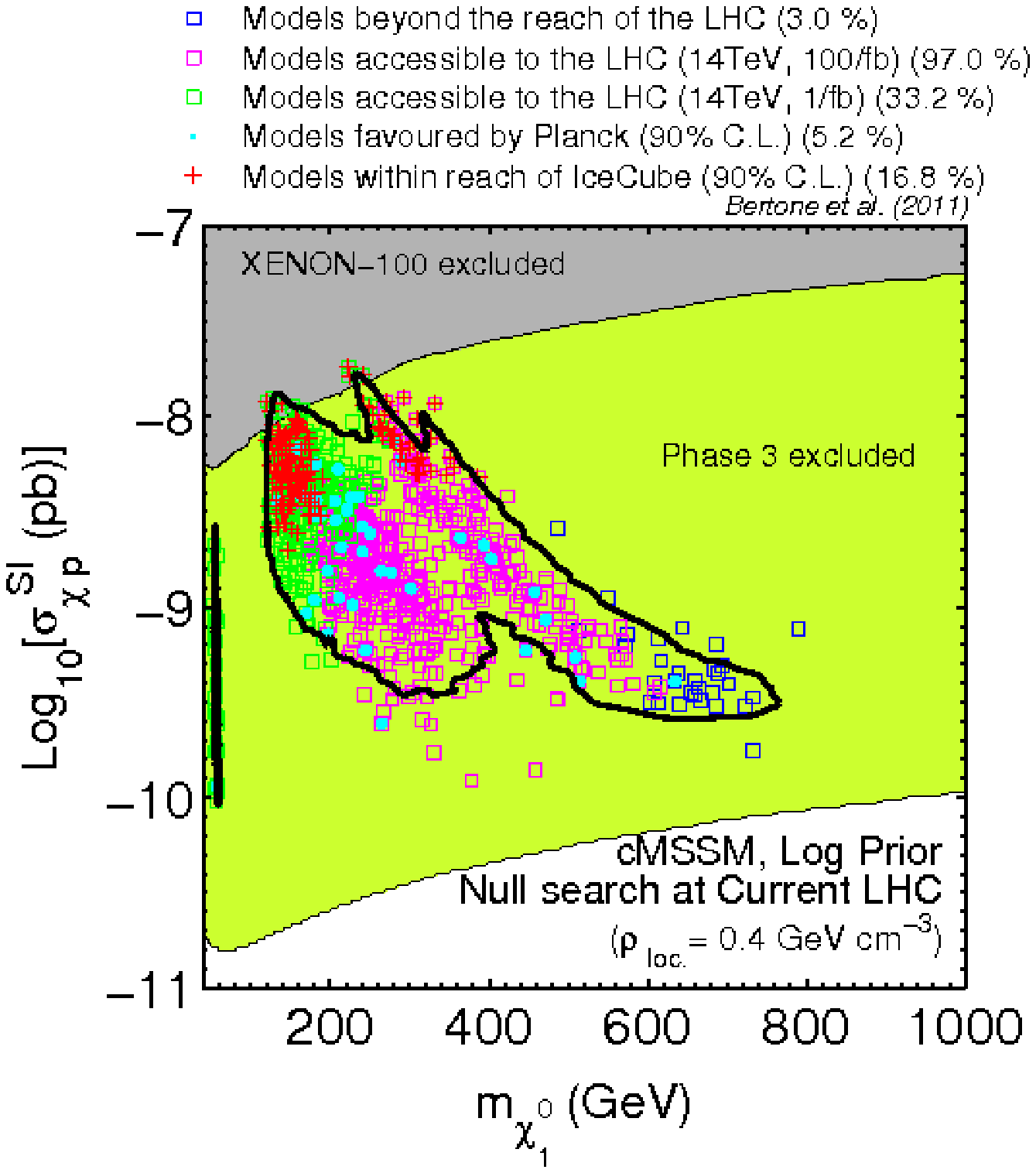}
\caption{Summary of the reach of various probes in the $\mchi$ vs $\sigsip$ plane 
for flat priors (left) and log priors (right) for samples resulting from a null search 
using the current configuration of the LHC. The solid, black contour delimits the 
95\% posterior region from current data (including Current LHC constraints). Equal 
weight samples from the posterior are shown and are coloured (and thinned by 
a factor of two for display purposes) as follows: green (magenta) squares are 
accessible to the LHC with $\sqrt{s}=14\,{\rm TeV}$ and IL\,=\,1\,fb$^{-1}$ 
(IL\,=\,100\,fb$^{-1}$); blue squares are outside the reach of the LHC with 
$\sqrt{s}=14\,{\rm TeV}$ and IL\,=\,100\,fb$^{-1}$ (nightmare scenario); cyan filled 
circles are expected to be favoured at 90\%\,C.L. by Planck; red crosses are accessible 
to IceCube. The yellow shaded region represents the reach of future Phase 3 direct 
detection experiments.}
\label{fig:summary}
\end{figure*}
\end{center}
 %***********************************************************************************

The capture rate is computed using \texttt{DarkSUSY v5.05} assuming, as above, 
a local CDM density $\rholoc=0.4\,$GeV\,cm$^{-3}$ and a Maxwell-Boltzmann 
velocity distribution with mean $\bar{v}=235\,$km\,s$^{-1}$. In order to assess 
the prospects for discovering DM with IceCube, the expected signal, given by 
Eq.\,(\ref{eq:Nmu}), must be compared with the the expected number of background 
events arising from the following sources:
 %***********************************************************************************
\begin{enumerate}[i)]
\item down-going muon events produced in a single cosmic ray shower in the southern hemisphere; 
\item muon events, misreconstructed as upgoing, resulting from two simultaneous cosmic ray showers in the southern hemisphere;  
\item an isotropic background of neutrino--induced muon events generated by atmospheric neutrinos; 
\item neutrino--induced muon events generated by cosmic neutrinos;
\item and muon events generated by neutrinos produced by cosmic-ray interactions in the solar corona.
\end{enumerate}
 %***********************************************************************************

We follow a methodology similar to that adopted in \cite{Albuquerque:2010bt} 
and we only consider events from the direction of the Sun when it resides in the 
northern hemisphere, thereby eliminating contribution from (i). The contribution 
from (ii) can be reduced to negligible levels by using appropriate analysis cuts 
and therefore it is reasonable to assume that IceCube can reject such a background
\cite{Carlos_priv_com}. The isotropic background contribution from cosmic 
neutrinos (i.e., background (iv)) can be safely neglected (we estimate it to be 
approximately $5\times10^{-3}$ events per year), as well as the background 
contribution (v) from cosmic-ray interactions in the solar corona, which we calculate
to be approximately $5\times10^{-4}$ events per year. Our estimate of background 
due to up-going atmospheric neutrinos (iii) is obtained by substituting into 
Eq.\,(\ref{eq:Nmu}) the best-fit measurements for the isotropic atmospheric 
neutrino flux from SuperKamiokande and AMANDA\,II presented in Fig.\,10 of 
\cite{Abbasi:2009nfa}, leading to an expected background, $\langle n_b\rangle$, 
of approximately 4.9 events per year. From this background, the number of muon 
events corresponding to a $90\%$\,C.L. detection threshold can be estimated 
using the Feldman-Cousins construction \cite{Feldman:1998}, giving $N_{90} = 32$ 
events for a 5 year period.

In Fig.\,\ref{fig:mchimuon} we display the posterior probability distribution of the 
cMSSM, in the ($\mchi$, N$_{\mu}^{\rm IC}$) plane, corresponding to null searches 
by the LHC with combinations of $\sqrt{s}$ and integrated luminosities as in 
Figs.\,\ref{fig:m0mhalf} and \ref{fig:mchisigsip_ansatz} for flat (upper) and log (lower) 
priors on cMSSM input parameters. We also display the corresponding number of 
events associated with the $90\%$\,C.L. Feldman-Cousins sensitivity estimate 
$N_{90}$ (dashed line). In Table\,\ref{tab:survival} we list the percentage of points 
inaccessible to the LHC that are detectable by IceCube (i.e., above the $N_{90}$ 
threshold) for each of our investigated LHC configurations. The left panels of 
Fig.\,\ref{fig:mchimuon} correspond to the current LHC configuration. For our log 
prior we can clearly identify the $h$-pole, stau co--annihilation and funnel regions 
as the 68\% C.L. regions spanning the mass ranges $50\,{\rm GeV}\lesssim\mchi\lesssim60\,{\rm GeV}$, 
$100\,{\rm GeV}\lesssim\mchi\lesssim500\,{\rm GeV}$ and $200\,{\rm GeV}\lesssim\mchi\lesssim1\,{\rm TeV}$
respectively. The distribution of probability in these regions is extremely similar to 
that displayed in Fig.\,\ref{fig:mchisigsip_ansatz} for the ($\mchi, \sigsip$) plane. 
Despite the fact that each of these three regions are partially accessible to IceCube,
only up to $\sim17$\% of the samples currently outside the reach of the LHC can 
actually be probed. For flat priors, both the stau co--annihilation and funnel regions 
are partially accessible (the $h$-pole region is barely visible due to the resolution 
of the scan), but the relative suppression of the stau co--annihilation region with 
respect to the log priors leads to an even smaller percentage, $\sim10\%$, of detectable 
samples. 
 %***********************************************************************************
\begin{center}
\begin{figure*}[t]
\includegraphics[width=0.32\textwidth,keepaspectratio, clip]{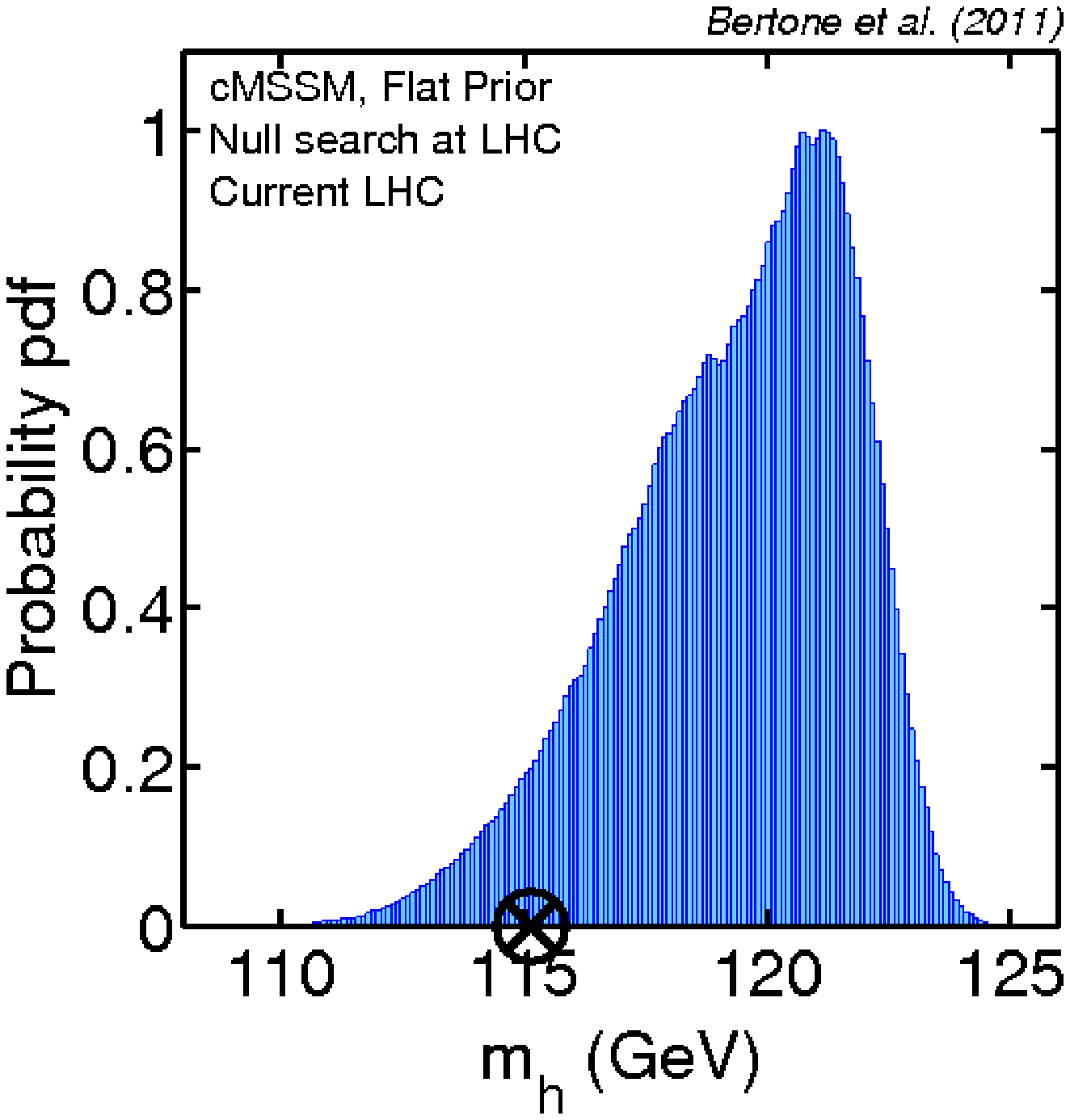}
\includegraphics[width=0.32\textwidth,keepaspectratio, clip]{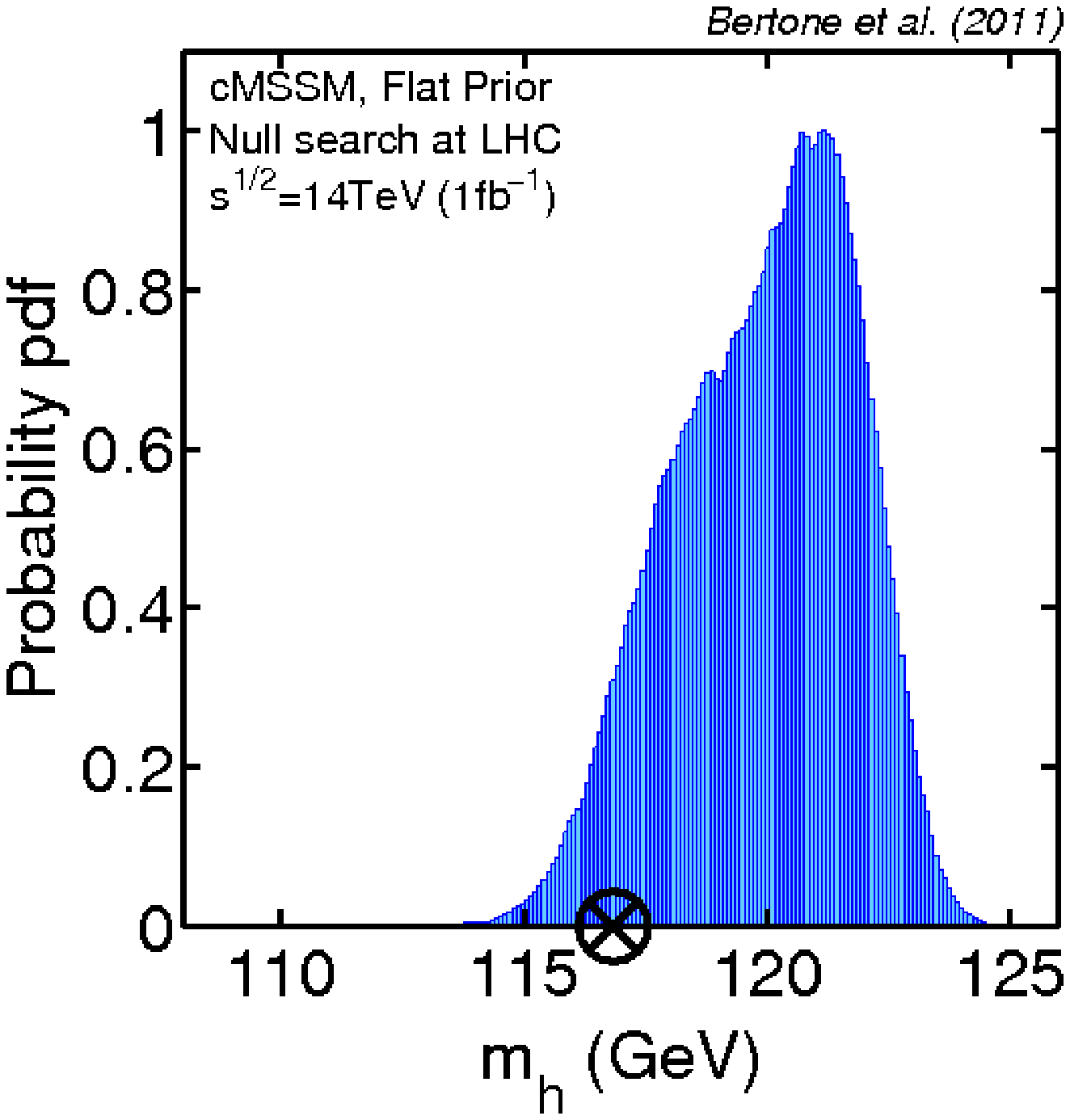}
\includegraphics[width=0.32\textwidth,keepaspectratio, clip]{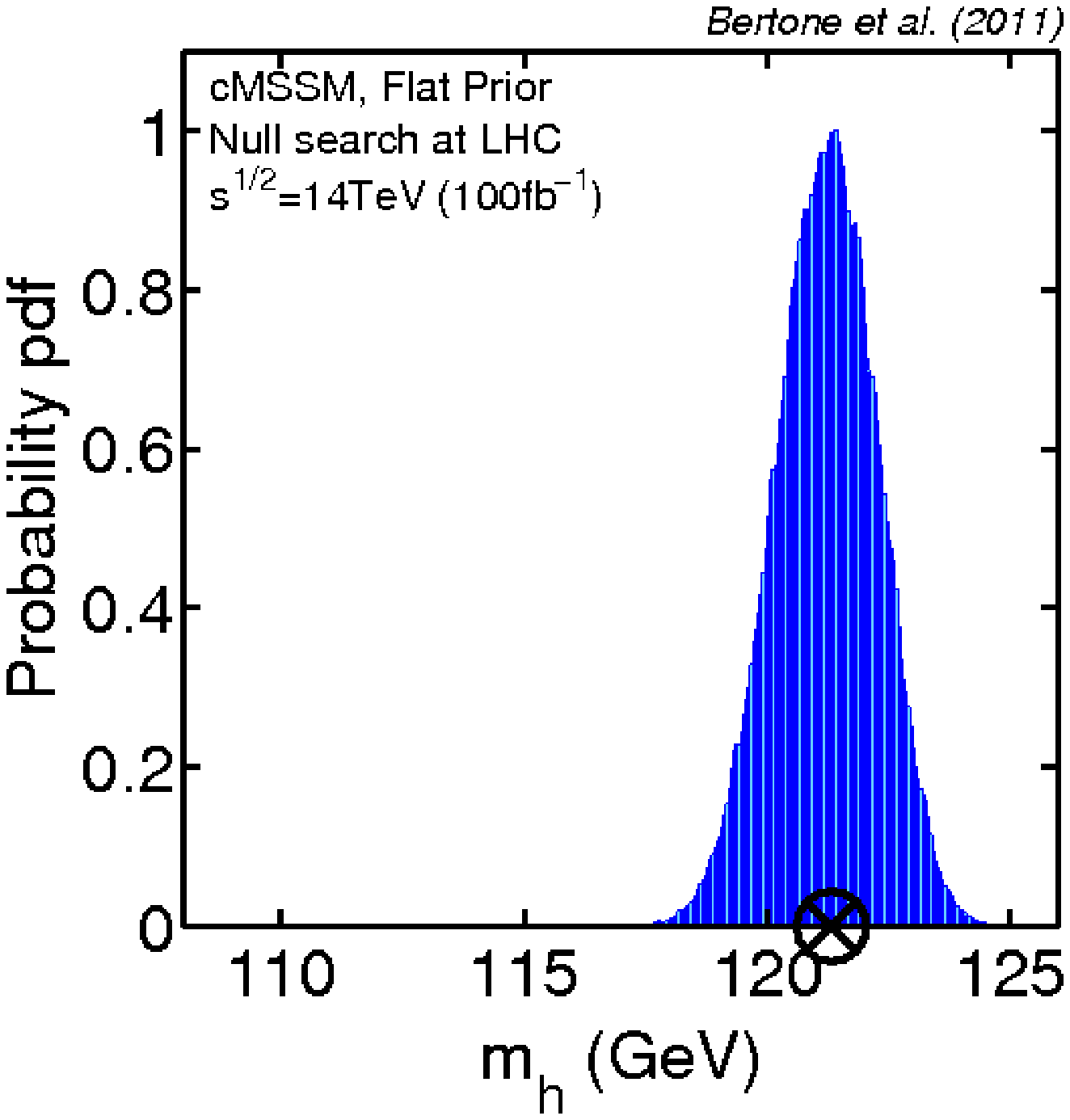}\\
\includegraphics[width=0.32\textwidth,keepaspectratio, clip]{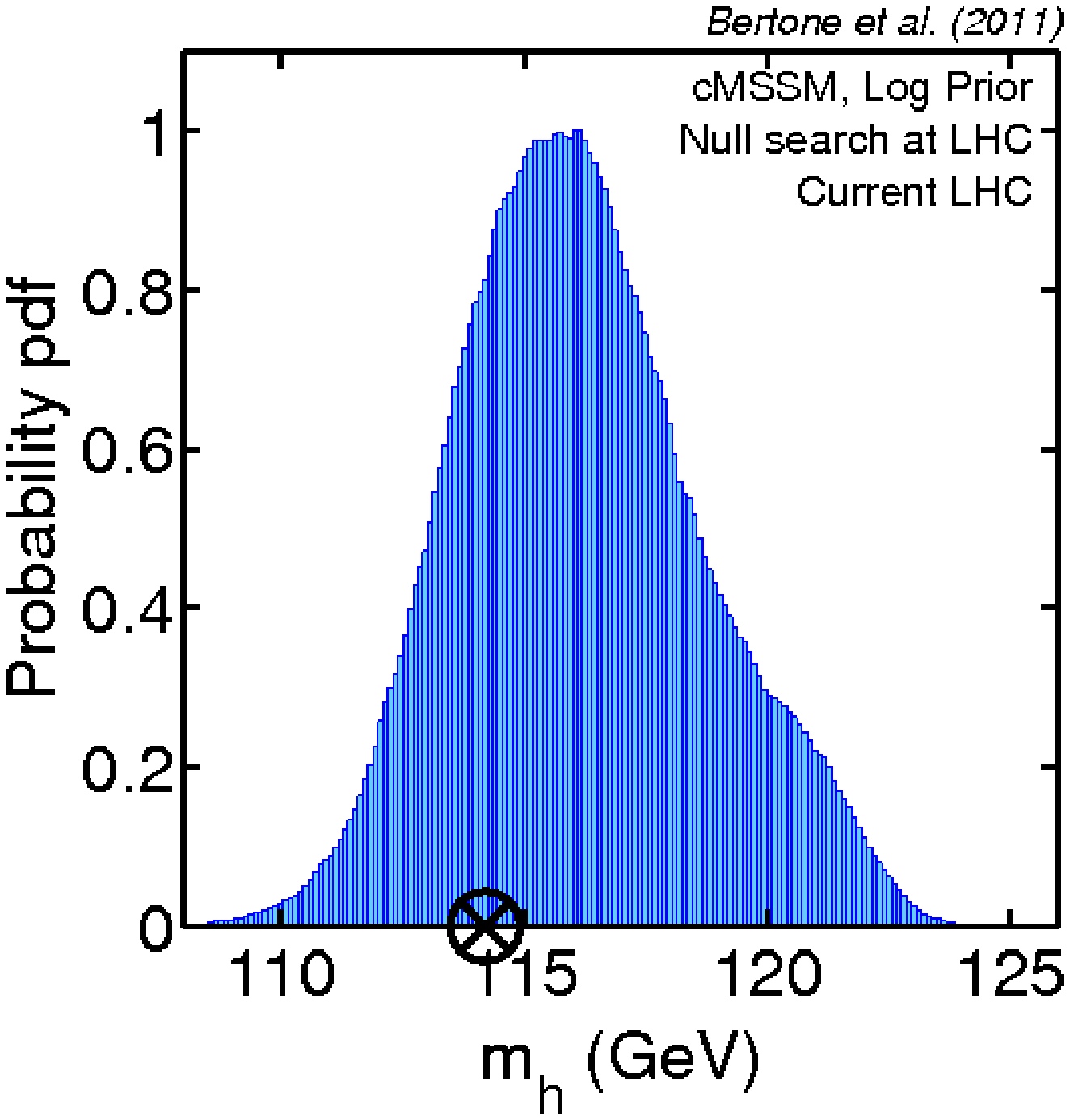}
\includegraphics[width=0.32\textwidth,keepaspectratio, clip]{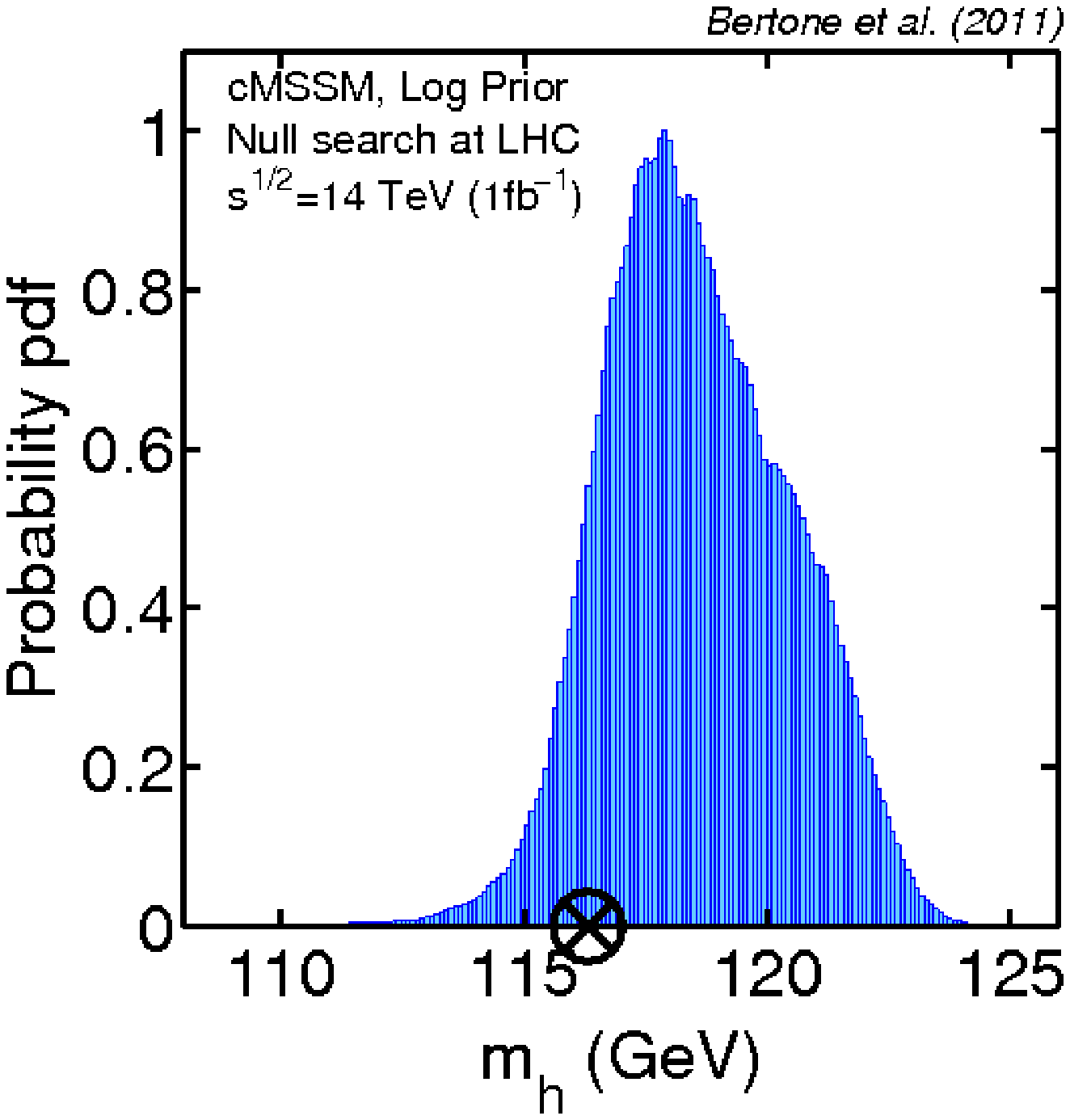}
\includegraphics[width=0.32\textwidth,keepaspectratio, clip]{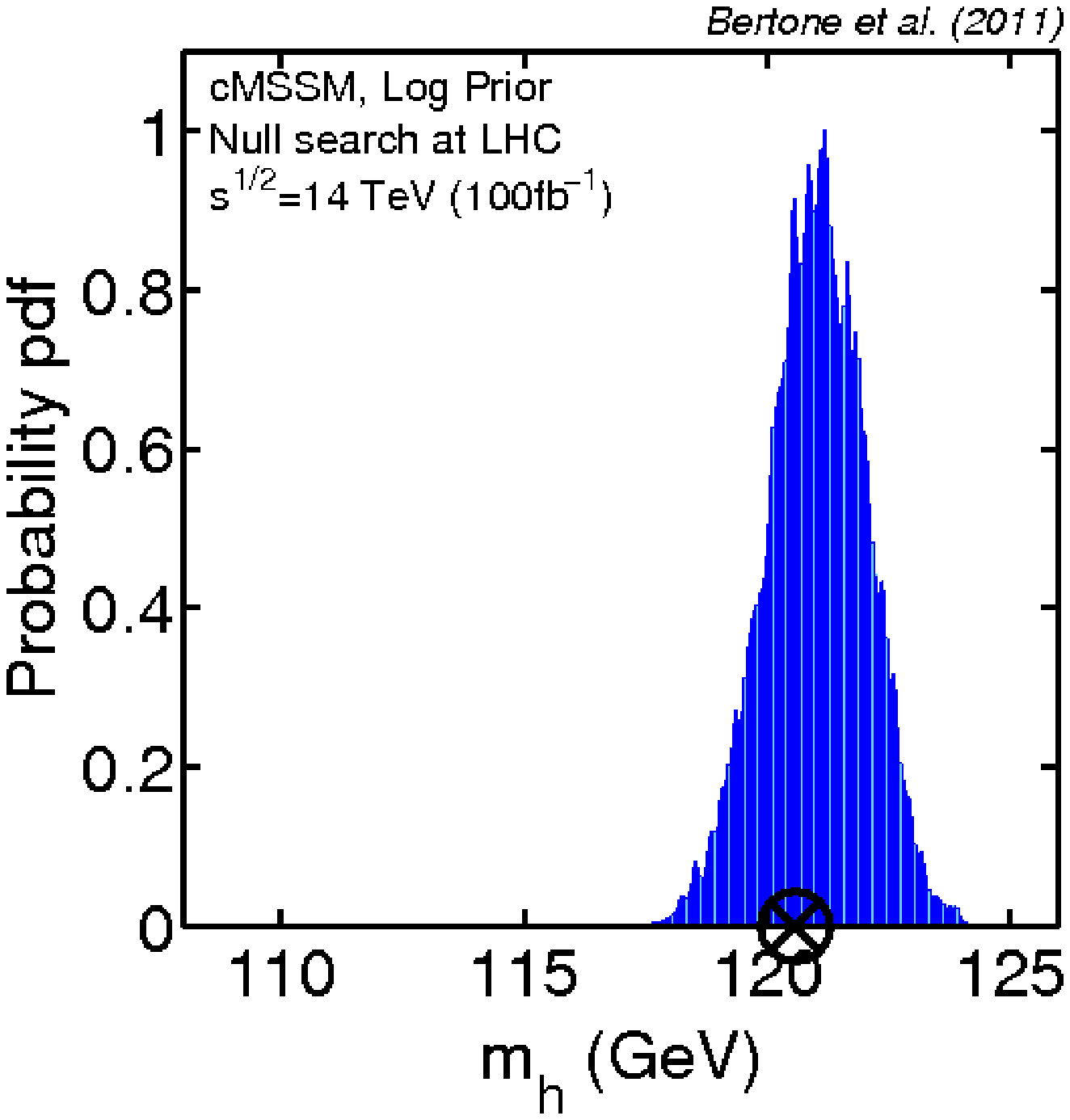}\\
\caption{1D pdf for the lightest Higgs mass $m_h$ for flat priors (top) and log priors 
(bottom), assuming Current LHC constraints (left), null searches at the LHC with 
$\sqrt{s}=$14\,TeV and IL\,=\,1\,fb$^{-1}$ (middle) and null searches at the LHC 
with $\sqrt{s}=$14\,TeV and IL\,=\,100\,fb$^{-1}$ (right panels). The encircled black 
cross represents the best--fit point. \label{fig:mh}}
\end{figure*}
\end{center}
 %***********************************************************************************

The central panels of Fig.\,\ref{fig:mchimuon} correspond to $\sqrt{s}=14\,{\rm TeV}$ 
and IL\,=\,1\,fb$^{-1}$. In analogy with the corresponding results in Fig.\,\ref{fig:mchisigsip_ansatz}, 
the $h$-pole region and a large portion of the stau co--annihilation region are probed 
by the LHC, and therefore ruled out in case of null searches. This leads to a shift 
of the posterior to higher neutralino masses, and to an enhancement in correspondence 
of the funnel region, which tends to reside for linear priors at $\mchi\sim500$\,GeV 
and $N_{\mu}^{\rm IC}\lesssim10$, hence inaccessible to IceCube. For log priors, 
the neutralino mass is shifted towards lower values, and $N_{\mu}^{\rm IC}$ towards 
slightly higher values, leading to a larger probability of detection, 11\%, to be compared 
with 6.9\% in the case of flat priors. 

Finally, for $\sqrt{s}=14\,$TeV and IL\,=\,100\,fb$^{-1}$ (right panels), IceCube will 
have little impact in probing the cMSSM in case of null searches at the LHC, for 
either sets of priors, resulting in detection probabilities in the range $0.1-1$\%.

We have also verified that for all LHC configurations, the entire parameter space 
fraction accessible to IceCube will also be accessible to Phase 1 direct detection 
experiments (but not vice-versa). This means that IceCube on its own is not expected 
to be able to improve constraints with respect to even Phase 1 direct detectors. However, 
it will provide a valuable independent cross check in at least a subset of the parameter 
space. This is obviously important in terms of evaluating and reducing systematic 
effects, e.g., from astrophysical uncertainties, that would affect both direct and indirect 
detection limits.

Several features in Figs.\,\ref{fig:mchisigsip_ansatz} and \ref{fig:mchimuon} can be 
better understood upon inspection of the 1D pdf for $\mchi$, shown in Fig.\,\ref{fig:mchi} 
for the same LHC configurations and same priors. The $h$-pole, stau co--annihilation 
and funnel regions can be roughly identified in the left panels, and the progressive 
shift towards large $\mchi$ as the LHC cuts into the stau co--annihilation and funnel 
regions is apparent from the central and right panels. It is also clear from this figure 
that the prior dependency of the results decreases as more and more constraining 
data are accumulated (left to right), as expected.

Our results are summarized in Fig.\,\ref{fig:summary}, which shows in a more concise 
way the parameter space accessible to various observational channels in the 
($\mchi$, $\sigsip$) plane, for both choices of priors. For this figure, we have additionally 
investigated which fraction of the parameter space that can be potentially ruled out by a 
more accurate determination of the relic abundance by the Planck satellite, which is 
expected to tighten current cosmological limits on the DM relic density by a factor of 
$\sim10$ (see, e.g., \cite{Plancksigma}). We find that about 95\% of the cMSSM parameter 
space outside the reach of the LHC will be potentially ruled out by Planck (at the 90\% C.L.), 
independently of the assumed LHC configuration. This is because models with the 
``correct'' relic density (i.e., matching the current determination by WMAP within Planck 
sensitivity) are essentially uniformly spread throughout the ($m_0, m_{1/2}$) plane 
(see the cyan samples in Fig.\,\ref{fig:summary}). Therefore, Planck has almost uniform 
power in probing them, independently of the parameter space fraction excluded by the 
LHC. Fig.\,\ref{fig:summary} also clarifies that the reach of IceCube lies just below the 
current direct detection exclusion limits from XENON-100 (red crosses), while the LHC 
nightmare scenario (i.e., the region inaccessible to the LHC with $\sqrt{s}=14\,{\rm TeV}$ 
and IL\,=\,100\,fb$^{-1}$, indicated here by the blue squares) is confined to higher neutralino 
mass regions of the parameter space, corresponding to spin-independent cross sections 
in the range $\sigsip \sim 10^{-9} - 10^{-10}$ pb. It is again apparent that Phase 3 DD 
experiments will be able to probe the entire surviving cMSSM parameter space, including 
the nightmare region. As mentioned above, further inclusion of astrophysical and hadronic 
uncertainties in the analysis is unlikely to change this conclusion, as the reach of Phase 3 
DD experiments lies comfortably below the favoured parameter space. 

%%%%%%%%%%%%%%%%%%%%%%%%%%%%%%%%%%%%%%%
\subsection{The role of the Higgs}
\label{subsec:Higgs}
%%%%%%%%%%%%%%%%%%%%%%%%%%%%%%%%%%%%%%%

What is the role of the Higgs boson in this scenario? Obviously, a detection of the 
Higgs in the next few years would provide crucial information, but since we do not 
know if, and at which mass, it will be detected, we can only discuss what consequences 
the possible outcomes of Higgs searches would have on our scenario.

To this aim, we have calculated the 1D pdf for the lightest Higgs mass $m_h$, 
shown in Fig.\,\ref{fig:mh} for the same LHC configurations and same priors as 
in Figs.\,\ref{fig:mchisigsip_ansatz} and \ref{fig:mchimuon}. In the so-called 
decoupling limit, the lightest Higgs couplings to Standard Model particles are 
identical to those of the Standard Model Higgs \cite{Djouadi:2005gj}. This happens 
when $m_A \gg m_Z$, a condition that is fulfilled for large values of $m_{1/2}$. Upon 
inspection of Fig.\,\ref{fig:m0mhalf}, we see that the posterior is indeed pushed towards 
large values of $m_{1/2}$, especially in the right--most plots, corresponding to 
IL\,=\,100\,fb$^{-1}$ of data at $\sqrt{s}=$14\,TeV.

A recent analysis of the prospects for detecting the Standard Model Higgs boson 
at ATLAS \cite{Collaboration:2010dk}, has shown that IL\,=\,2 fb$^{-1}$ of data at $\sqrt{s}=$8\,TeV 
will be sufficient to exclude Standard Model Higgs masses between 114 and 500\,GeV 
at the 95\% CL. Since our 1D pdf's for null searches at the LHC with 14 TeV fall in the 
range 115--125 GeV, we can conclude that if the LHC will be able to rule out the Higgs in this range, then the cMSSM will be excluded. On the other hand, a $5\sigma$ discovery of the Higgs might take a much larger luminosity and might require running at larger energies. So if by the end of 2012 the LHC will have failed to rule out the last remaining window for the Higgs (currently, between 115 and 140 GeV, considering the recently presented results from the ATLAS and CMS combination of $1.6+1.6$ fb$^{-1}$ of data \cite{Rolandi2011}), then the future generation of direct detection experiment will have a good chance of discovering dark matter (within the cMSSM framework considered here).  

%%%%%%%%%%%%%%%%%%%%%%%%%%%%%%%%%%%%%%%
\section{Discussion and Conclusions}
\label{sec:summary}
%%%%%%%%%%%%%%%%%%%%%%%%%%%%%%%%%%%%%%%

In this study we have quantitatively assessed the potential for discovering the 
cMSSM in case of null searches at the LHC (the ``nightmare scenario" of particle 
physics), with future direct detection experiments and with the IceCube neutrino 
telescope.

Our main conclusion is that Phase 3 direct detection experiments (that can reach 
cross sections down to $\sigsip\sim10^{-11}$\,pb) will be able to probe entirely 
the favoured region of the cMSSM parameter space in the nightmare scenario of 
particle physics, therefore providing a unique opportunity to test SUSY even in 
case of null searches at the LHC. Interestingly, these experiments are expected 
to be built on a timescale of 5--10 years, similar to that required for the LHC to 
reach IL\,=\,100\,fb$^{-1}$ of data at $\sqrt{s}=$14\,TeV.

In our analysis, we fixed all astrophysical parameters to the standard values commonly 
adopted in the literature. In principle, however, one should take into account all particle 
physics and astrophysical uncertainties in order to combine in a self-consistent fashion 
accelerator and particle astrophysics experiments \cite{Bertone:2010rv,Pato:2010zk,
Serpico:2010ae}. Including the uncertainty on the local DM density \cite{Pato:2010yq,
Salucci:2010qr} would not change the fraction of cMSSM points accessible to direct 
detection experiments, since the SUSY points and the sensitivity curves would be 
rescaled by the same quantity (i.e., the ratio of the `true' local density over the value 
adopted here). A more careful treatment of the velocity distribution and of the hadronic 
uncertainties in the neutralino-nucleon cross-section may instead have a stronger 
impact on direct and indirect searches \cite{Ellis:2008hf, Strigari:2009zb,Serpico:2010ae},
 but given the ample margin between the bulk of the nightmare cMSSM parameter 
 space and the sensitivity of Phase 3 direct detection experiments, it is unlikely 
 that our main conclusion would change significantly. 
 
Finally we have studied the implications of Higgs searches at the LHC, and we 
have shown that if a Standard Model-like Higgs is not found at the LHC with IL\,=\,100\,fb$^{-1}$ 
of data at $\sqrt{s}=$14\,TeV, that would basically rule out the cMSSM, while an actual detection 
in the appropriate mass range would provide additional motivation to continue the 
study of Supersymmetry with astroparticle experiments.
 
%%%%%%%%%%%%%%%%%%%%%%%%%%%%%%%%%%%%%%%
\section*{Acknowledgments}
%%%%%%%%%%%%%%%%%%%%%%%%%%%%%%%%%%%%%%%
We would like to thank Carlos Perez de los Heros for his advice regarding IceCube 
and DeepCore and our co-authors of Ref.~\cite{Bertone:2011nj} for providing us 
with the MultiNest samples obtained in their work. DTC is supported by the Science 
and Technology Facilities Council. Part of this work was conducted while DTC 
was at CERCA Case Western Reserve University, supported by NASA grant 
4200188792 and also supported in part by the US DoE. R. RdA would like to thank 
the support of the Spanish MICINN's Consolider-Ingenio 2010 Programme under 
the grant MULTIDARK CSD2209-00064. R.T. would like to thank the Kavli Institute 
for Theoretical Physics, Santa Barbara, for hospitality. This research was  supported 
in part by the National Science Foundation under Grant No. PHY05-51164.

%%%%%%%%%%%%%%%%%%%%%%%%%%%%%%%%%%%%%%%

%%%%%%%%%%%%%%%%%%%%%%%%%%%%%%%%%%%%%%%

%%%%%%%%%%%%%%%%%%%%%%%%%%%%%%%%%%%%%%%
\end{document}